\documentclass[12pt]{iopart}
\usepackage{epsfig}
\usepackage{epstopdf}
\usepackage [latin1]{inputenc}
\usepackage{color}
\usepackage{enumerate}
\usepackage{dsfont}
\usepackage{cite}
\usepackage{mathrsfs}

\begin{document}
{\bf 
Can biological quantum networks solve NP-hard problems?
}
\author{G\"oran Wendin}
\address{Microtechnology and Nanoscience - MC2\\
Chalmers University of Technology\\
SE-41296 Gothenburg, Sweden\\
Email: goran.wendin@chalmers.se}

\date{\today}

\begin{abstract}

There is a widespread view that the human brain is so complex that it cannot be efficiently simulated by universal Turing machines. During the last decades the question has therefore been raised whether we need to consider quantum effects to explain the imagined cognitive power of a conscious mind. 

This paper presents a personal view of several fields of philosophy and computational neurobiology in an attempt to suggest a realistic picture of how the brain might work as a basis for perception, consciousness and cognition. The purpose is to be able to identify and evaluate instances where quantum effects might play a significant role in cognitive processes. 

Not surprisingly, the conclusion is that quantum-enhanced cognition and intelligence are very unlikely to be found in biological brains. Quantum effects may certainly influence the functionality of various components and signalling pathways at the molecular level in the brain network, like ion ports, synapses, sensors, and enzymes. This might evidently influence the functionality of some nodes and perhaps even the overall intelligence of the brain network, but hardly give it any dramatically enhanced functionality. So, the conclusion is that biological quantum networks can only approximately solve small instances of NP-hard problems.

On the other hand, artificial intelligence and machine learning implemented in complex dynamical systems based on genuine quantum networks can certainly be expected to show enhanced performance and quantum advantage compared with classical networks. Nevertheless, even quantum networks can only be expected to efficiently solve NP-hard problems approximately. In the end it is a question of precision - Nature is approximate.

\end{abstract}

\maketitle

\tableofcontents

\newpage


\section {Introduction}

Human consciousness is often considered to be a hard problem that cannot be treated efficiently by classical computers. To explain the assumed superior capacity of the brain, it is sometimes speculated that quantum coherence and entanglement may be part of the solution.

Despite the present lack of understanding of the physical mechanisms behind the information processing of the brain, there is however one thing to be noted - quantum computing does not solve NP-hard problems efficiently (see e.g. \cite{Aaronson2005,Aaronson2013,Broersma2018,Wendin2017}). A quantum processor in the brain could possibly accelerate certain cognitive abilities, but hardly provide any major contribution to explaining the "Hard problem of consciousness" \cite{Chalmers1995}.

A well-suppported current view is that the brain is a classical complex dynamical system showing self-organised critical behaviour  (SOC) \cite{Bak1996,Bullmore2009,Bassett2011,Chialvo2010,Tagliazucchi2016,Tagliazucchi2017,Lake2017,Cocchi2017,Maass2016}. In SOC models, the physical basis for consciousness may be emergent long-range functional correlations. Obviously, local quantum coherence and entanglement in specific brain and sensor components might influence the details of the complex dynamics. However, this will hardly enhance the computational power of the brain. And how that power is related to human perception, consciousness and thinking is probably a very different question. 

The core of the problem is that quantum physics is often conjectured or postulated to add power to cognition without knowledge of what cognition really is. 
It certainly seems reasonable to assume that cognition is connected with physical processes in the brain, and quantum effects could possibly affect such processes. Whether that would influence  consciousness, cognition and subjective experience is a different matter. 
In order to discuss quantum enhanced cognition, we need to relate to concrete classical computational models of the brain that effectively have not yet been formulated. However, even if they were known, the answer to the question of "Can biological quantum networks solve NP-hard problems?" would be a clear personal "no". Not even quantum computers can efficiently solve NP-hard problems.

It is extremely interesting to see how the topic of consciousness and cognition in neuroscience and philosophy, with or without quantum effects, has evolved during the last three decades through the opinions expressed in the work of professionals in neuroscience, psychiatry, psychology, physics and philosophy that forcefully defend their established lines of thought and structures of knowledge \cite{ChangeuxDehaene1989,Dehaene2001,DehaeneChangeux2011,Dehaene2014,Dehaene2017,Barttfeld2015,KochHepp2006,Tononi2004,Tononi2008,Casali2013,Tononi2016,Koch2017,Koch2018,Dennett1991,Dennett2017,Dennett2018,Nagel1974,Nagel2012,Nagel2017}. On the whole, there is little convergence. It is difficult to avoid the feeling(!) that the good old dividing line is more visible than ever: materialists versus dualists.

Only recently becoming conscious of those paradigms and "schools", I tend to agree with Dennett \cite{Dennett2018} that the canonical "Hard problem of consciousness" of Chalmers \cite{Chalmers1995} is a chimera, leading us astray.  This is also the view of Dehaene \cite{Dehaene2014}.
To me, the origins of consciousness and cognition are very concrete things that are already basically understood from a systems point of view of the brain. How that connects to subjective experience and personal perceptions - qualia - is a question of internal representation and reporting for human individuals. "Intentionality, meaning, and value" are certainly valid abstract concepts, but are necessarily coded and processed as physical information.

The mission of this paper is threefold: (i) to put quantum physics in its proper place biologically, (ii) to make probable that there are relevant top-down computational classical models of the brain meeting established bottom up experimental neurophysiological models, and (iii) to clarify where quantum effects just might play a role for cognition. No need to invent a universal mind, just hard persistent work to convince Nagel \cite{Nagel2012} that the modern materialist approach to life will {\it not fail} to explain the physical/biological foundations for consciousness, intentionality, meaning, and value. What is then the meaning and value of existence is a very different matter.

First we outline some views on the power of computation (Sect. 2), present some philosophical background (Sect. 3), and describe coupled  brain networks as dynamical complex systems (Sect. 4).  Section 5 presents a review and analysis of the status of experimental knowledge of the structure and dynamics of global brain networks. The quantum aspects of cognition are discussed in Sect. 6, followed by some comments on quantum networks in Sect. 7.

\newpage
  
\section {What is the power of  computing?}

Unconventional  computing (UCOMP) is a vast field covering various kinds of information processing different from von Neumann type of classical digital computing \cite{Broersma2018,CompMatter2018}. The field deals with various forms of analogue computing, and key questions are \cite{Broersma2018}:
(i) Can UCOMP provide solutions to classically undecidable problems?
(ii) Can UCOMP provide more effective solutions to NP-complete and NP-hard problems?
(iii) Are classical complexity classes and measures appropriate to any forms of UCOMP?
(iv) Which forms of UCOMP are clearly and easily amenable to characterisation and analysis by these? And why?
(v) Are there forms of UCOMP where traditional complexity classes and measures are not appropriate, and what alternatives are then available? 

There is a long-held notion that some forms of unconventional classical computing can provide tractable solutions to NP-hard problems that take exponential time or memory resources  for classical digital machines to solve \cite{Adleman1994,Siegelmann1995,SiegelmannSonntag1995,Cabessa2011}, and the idea of solving NP-hard problems in polynomial time with finite resources is still actively explored \cite{DiVentraTraversa2018,Pei2017,Manukian2018,Sheldon2018,Traversa2018}. 
The question is then, what is the real computational power of UCOMP machines: are there UCOMP solutions that can at least provide significant polynomial speed-up and energy savings, or more cost-effective solutions?

The brain can be modelled as a heterogenous adaptive recurrent neural network (RNN). RNNs with rational weights are Turing-complete, meaning that they can simulate a universal Turing machine and vice versa. If the weights represented real numbers, the RNN would be super-Turing. However, this requires infinite precision and is not a realistic situation. The best we can achieve is most probably the computational power of adaptive RNNs \cite{Sussillo2014,Carmantini2015}.

Since analogue machines like  RNNs can be efficiently simulated by digital computers \cite{Vergis1986} with "only" polynomial overhead, the conclusions of several decades of discussion and debate \cite{Aaronson2005,Aaronson2013,Denef2007,Douglas2013} can be illustrated in terms of a number of "Laws"  \cite{Broersma2018}:
1st Law of Computing: You cannot solve uncomputable or NP-hard problems efficiently unless you have a physical infinity or an efficient oracle.
2nd Law of Computing: There are no physical infinities or efficient oracles.
3rd Law of Computing: Nature is physical and does not solve uncomputable or NP-hard problems efficiently.
Corollary: Nature necessarily solves uncomputable or NP-hard problems only approximately. 

A natural consequence of the Corollary is the following: Natural problems solved by the Evolution do not represent really hard instances. Protein folding is in principle NP-hard, scaling exponentially with time. The protein chains chosen by Nature to build biological systems must therefore involve fairly "simple" instances of NP-problems - otherwise we would not exist, folding would take too long. However, protein folding in biological systems does not take place from scratch - testing all instances and configurations. Nature is adaptive and coevolution has created efficient processes for assembling complicated proteins on catalysing enzymatic  substrates (chaperones) in finite short time. 

It follows that if we simulate a such a process digitally we will need adaptive machine-learning algorithms that can find the optimal routes through the energy landscape. In principle it also implies that it would be more efficient to use physical implementations of RNNs. 
The problem is that we cannot build physical RNNs with sufficient complexity - so far it has been necessary to simulate them in software using powerful digital computers.

This is now beginning to change by the recent technological development (IBM, TrueNorth \cite{Merolla2014}), making it possible to emulate spiking neural networks directly in state-of-the-art semiconductor hardware \cite{Merolla2014,Esser2016,Maass2016a}, building exremely fast event-based systems for image processing \cite{Esser2016,Amir2017,Andreopoulos2018} and deep learning \cite{Ambrogio2018}. 

These conclusions also apply to quantum information processing (QIP). Certainly QIP is able to efficiently compute many problems that will take exponential resources for classical computers. But in the end, not even QIP can solve NP-hard problems efficiently.
It follows that quantum networks cannot solve NP-hard problems efficiently - no good reason for making the brain a quantum processor in order to explain its computational capacity.

\newpage

\section {What do philosophers say about mind and matter?}

The question of the nature of consciousness and self-awareness has bothered philosophers from little Lucy until today.
The litmus-test is supposedly how you react when you look into a mirror -  always a troublesome experience. 
Over the years, the issue of consciousness -  especially self-consciousness - was mostly addressed by philosophers and theologists.
During the 1960s a philosophical materialist view - Functionalism \cite{Churchland2004} - was developed, assuming that the brain is a neuronal machine, and that everything can be explained on that basis. The quality of self-consciousness and the human brain has then come into new light with the advent of powerful computers and "deep-learning" adaptive algorithms, as well as through the quite dramatic progress in dynamic brain imaging.

The basic problem is the lack of operational definition and measure of consciousness and qualia. Philosophers have tended to group into opposing categories of "materialists" and "dualists".  Materialists assume that consciousness is governed by physical law and can be understood in physical and chemical terms. They start from generic brain network models, investigate the dynamics, and compare with experimental results of brain imaging.
Dualists argue that mind and brain are radically different kinds of thing -  the mind cannot be treated as simply part of the physical world \cite{SEP2016}. The dualist perspective is expressed by David Chalmers in terms of the "Hard problem of consciousness"  \cite{Chalmers1995}. Chalmers claims that the problem of subjective experience - qualia - is distinct from the set of physically computable brain processes and that the problem of subjective experience will "persist even when the performance of all the relevant functions is explained". 

Dualists tend to invoke concepts that look supernatural from a materialist point of view. And, so far, materialists have not been able to explain the physical nature of consciousness and perception based on the workings of neural circuits and systems.
This has led to a "battle between philosophers", illustrated by a series of papers and exchanges over the last two decades \cite{Chalmers1995,Dennett1991,Dennett2017,Nagel2012,Nagel2017,Dennett1996,Chorost2013,Papineau2017}.
The essence is illustrated by Nagel's conclusion \cite{Nagel2012} that the modern materialist approach to life has failed to explain consciousness, intentionality, meaning, and value. Nagel  states that "this failure threatens the entire naturalistic world picture, extending to biology, evolutionary theory, and cosmology", underpinning the dualist view that it is impossible to reduce the mental level to the physical level. 

An opposing view is that it rather represents a failure to recognise a basic fact of physics, formulated by the late Rolf Landauer: {\it information is physical} \cite{Landauer1991}. "Intentionality, meaning, and value" are certainly abstract concepts, but they are also consequences of human cognition and thinking - they don't exist in free space independent of brains thinking about them. To have an opinion about the meaning or value of something takes some thinking and costs energy. Subjective experience - qualia -  necessarily involves information, and information is {\it not abstract} -  it has substance and it involves energy and entropy. Transfer of information - communication - takes power.

This was succinctly expressed in 2014 by Dehaene \cite{Dehaene2014}: Chalmers' hard problem of representing subjective experience is an ill-posed problem - the really  {\it hard} problem is the complex of computationally hard problems describing the intricate dynamics of complex brain networks. Chalmers' hard problem will dissolve when computational modelling and brain imaging advances.

We might actually be there already: what is the neural representation of subjective experience when a dose of the psychedelic psilocybin dissolves a person's entire self, traceable to changes in specific synaptic couplings of specific brain networks, disconnecting the individual from previous "programming" \cite{Carhart-Harris2014,Carhart-Harris2017a,Carhart-Harris2017b,Fontes2015}?

Arguably these changes of consciousness and self-awareness must correlate with changes of the intrinsic dynamic states of the network. These can be very sensitive to functional network connectivity. What is considered as "the self" in the absence of the drug, might be regarded as illusions (or the true self!) in the presence of the drug, or schizophrenic delusions in the case of pathological synaptic imbalance. 
A radical approach is to regard consciousness and self-awareness as illusions by default, like Dennett does \cite{Dennett2017,Dennett2018}, but this feels like a philosophical impasse.

\newpage

\section {The brain as a dynamic physical system}

\subsection {Large-scale brain networks}

The brain forms a complex dynamic system \cite{McKenna1994} that can be modelled as a highly clustered small-world type of network \cite{WattsStrogatz1998} of highly inter-connected modules built from networks of spiking neurons. These modules form nodes that are sparsely connected to each other \cite{Meunier2010,BresslerMenon2010,Menon2011,Andrews-Hanna_2012,Andrews-Hanna_2014}, allowing both local and global information processing on a variety of timescales \cite{Maass2016,Dehaene2017,Meunier2010}. 

Multiscale modelling of the structure and non-linear dynamics of the brain networks is a vast and active research field \cite{Bullmore2009,Bassett2011,Chialvo2010,Maass2016,ChangeuxDehaene1989,Dehaene2001,DehaeneChangeux2011,Dehaene2014,Dehaene2017,Meunier2010,Lee2017,Baars1988,Baars2002,Baars2005,Breakspear2017,Olbrich2011,Werner2007,Kitzbichler2009,Kitzbichler2015,BeggsTimme2012,Zhou2015,Papo2017,Racz2018,Werner2010,Newman2006,Enel2016,Rodriguez2017,Taylor2017,Grossberg2017,Seguin2018,vandenHeuvel2014,Wagner2017,Witter2015,Tognoli2014,Rossert2015,Stepp2015}, providing a computational foundation for understanding information processing in the brain and the emergence of consciousness and cognition. 

To be able to critically assess the potential role of quantum effects on consciousness and cognition, we must first outline an up-to-date picture of the brain's network structure, couplings and dynamics. 
Spontaneous brain activity is organised into resting-state networks (RSNs) involved in internally guided, higher-order mental functions (default mode (DMN), central executive (CEN), and salience (SN) networks \cite{Menon2011}), and externally-driven, specialised motor and sensory processing  (sensorimotor (SMN),  auditory (AN), and visual (VN) networks) \cite{Lee2017}. 
Cerebellar (CN) networks lie in between, involving both mental and motor processing.

The Default (Mode) Network (DMN)  \cite{Andrews-Hanna_2012,Andrews-Hanna_2014} describes the resting state function of the brain and is responsible for {\it intrinsic awareness} and {\it self-generated thought}. The Central Executive Network (CEN) (also called Executive Control Network, ECN \cite{Qin2015}) is responsible for extrinsic awareness (focused attention) that regulates
executive functions that control and mediate cognitive processes, including working memory, reasoning,
flexibility, problem solving, and planning. The Salience Network (SN) \cite{Qin2015} is implicated in the detection and integration of emotional and sensory stimuli, and in modulating the switch between the DMN and CEN, i.e. switching between introspection and attention. The connectivity within the SN distinguishes between unconscious and minimally conscious states \cite{Qin2015}. The Frontal Mesocircuit Network (FMN) \cite{Lant2016} serves as the basic underlying dynamic driver of the forebrain cortical networks and is necessary for supporting consciousness and governing the quality of consciousness. The  Cerebellar Network (CN) \cite{Marek2018} is of central importance for motor control, and is involved in some cognitive functions such as attention and language as well as in regulating fear and pleasure responses. 
In addition, there is a number of more local specialised attention/action networks for input/output functions: the Sensorimotor Network (SMN) \cite{Martino2016}, the Auditory network (AN), the Visual Networks (VN), and the Olfactory Network  (ON).
Disruptions in activity in various networks have been implicated in neuropsychiatric disorders such as depression, Alzheimer's, autism spectrum disorder, schizophrenia and bipolar disorder (see e.g. \cite{Menon2011,Berman2016,Lant2016,Martino2016}).

\subsection {Brain networks show self-organised  critical dynamics}

There is a host of evidence that much of the brain dynamics is poised at the edge of chaos \cite{Maass2016,Kitzbichler2009,Kitzbichler2015,BeggsTimme2012,Zhou2015,Papo2017,Racz2018}. Three decades ago, Kauffman  \cite{Kauffman1991} described criticality and coevolution to the edge of chaos and avalanches in biological systems, and Bak \cite{Bak1996} introduced the concept of self-organised criticality (SOC) and illustrated the fractal structure of avalanches at all scales with the formation and behaviour of a pile of sand. SOC represents a "global" systems description in terms of universality and critical exponents, and applications to real systems for interpretation of the dynamics must be done with great care, as is evident from a history of successes, controversies and open questions \cite{Werner2007,Kitzbichler2015,BeggsTimme2012,Watkins2016,Roli2018}.

\subsection {The brain performs reservoir computing}

As already discussed, there is now plenty of experimental support for modelling the conscious brain as a complex dynamic system poised at the edge of critical behaviour, separating chaotic and ordered states  \cite{Chialvo2010,Tagliazucchi2016,Tagliazucchi2017,Lake2017}.
In SOC models, the physical basis for consciousness may be emergent long-range functional correlations. 
In this situation the system is very sensitive to external perturbations, like input from different kinds of sensors, and it can produce fast response. 

Reservoir computing (RC) \cite{Maass2016,Schrauwen2008,LukoseviciusJaeger2009,Maass2009,Yamazaki2007,Konkoli2018a,Konkoli2018b,Inubushi2017} is a brain-inspired machine learning framework \cite{Maass2016}  that employs a signal-driven dynamical system, in particular harnessing common-signal-induced synchronisation which is a widely observed nonlinear phenomenon. 
RC has been applied to a variety of physical systems \cite{Sillin2013,Demis2015,VanderSande2017,Du2017,Pecqueur2018},
and in particular to information processing in brain-like networks \cite{Enel2016,Rodriguez2017,Yamazaki2007,Inubushi2017}. 
In particular, Enel {\it et al.} \cite{Enel2016} compared reservoir model activity to neural activity of the dorsal anterior cingulate cortex of monkeys, training a reservoir to perform a complex cognitive task initially developed for monkeys. The reservoir model inherently displayed a dynamic form of mixed selectivity \cite{Fusi2016}, key to the representation of the behavioral context over time, and  revealed similar network dynamics as the cortex \cite{Enel2016}.

A comprehensive and very readable discussion of the present status of brain reservoir computation is provided by Maass  \cite{Maass2016}. Maass emphasises that neural networks in the brain are not the kind of homogeneous artificial neural networks that are suitable for generic computation of Boolean circuits, or searches in energy landscapes for optimisation of cost functions.
Neural networks in the brain consist of feed-forward and feed-back coupled heterogeneous nodes containing a variety of spiking neurons that are connected by different types of synapses with short-term and long-term dynamics. The short-time dynamics of excitatory and inhibitory neurons/synapses depend on the structure of the pulse trains, making it history-dependent. All of this puts important constraints on computational models.

One might think that such a heterogeneous recurrent neural network might not be suitable for computations. However, Nature has successfully evolved such brains, and they are sometimes regarded as superior to digital computers. Naturally they may therefore be difficult to simulate. Nevertheless, there is a computational model for which a diversity of computational units and time constants is a clear advantage, namely the liquid computing paradigm \cite{Maass2016}, which together with the "echo-state" model \cite{Jaeger2001} is referred to as reservoir computing (RC). The liquid computing model is designed for biological neural networks governed by noise, diversity of units, and temporal aspects related to spikes.

Following Maass \cite{Maass2016}, RC models conceptually divide neural network computations into two stages, a fixed generic nonlinear pre-processing stage and a subsequent stage of linear readouts that are trained for specific computational tasks. In the pre-processing stage, different types of neurons and subcircuits within the recurrent network produce a large number of potentially useful features and nonlinear combinations of such features, out if which projection/readout neurons can select useful information for its target networks. In this way even seemingly chaotic dynamics of a recurrent local network can make a useful computational contribution. Furthermore, there is a connection to  machine learning using Support Vector Machines (SVMs). An SVM also consists of two stages: a generic nonlinear preprocessing stage (kernel) and linear readouts. The kernel projects external input vectors nonlinearly onto vectors in a much higher dimensional space. One can view a large nonlinear recurrent neural network as an implementation of such a kernel, where the network
response to an input corresponds to the kernel output \cite{Maass2016}. An important advantage of a two-stage computational model is multiplexing: the same first stage (the kernel) can be accessed by an unlimited number of linear readout neurons that learn to extract different computational results for their specific target networks.

Experiments show that the networks in the brain are changing continuously: connectivity, neurotransmitters and neural codes. Nevertheless, those spiking networks provide stable computational function in spite of ongoing rewiring and network perturbations  \cite{Maass2016}. Probabilistic computations are now playing a prominent role in many models in neuroscience. Stochasticity of spiking neurons enables the network to solve problems in a heuristic manner - each network state represents a possible solution to a problem, and the frequency of being in this network state encodes the fitness of the solution. This suggests a possible relation between flickering internal states of brain networks, and the response in terms of perception and behaviour.

\subsection{Brain network percolation model of sensory transmission}

Current neurocomputational efforts often state that the goal is to understand consciousness and cognition manifested in sufficiently complex  biological networks. 
In order to do precisely that, in a recent theoretical study of a model for thalamo-cortical structure, Zhou {\it et al.} \cite{Zhou2015} described a systems-level network approach to explaining brain network processing under deep anesthesia, to account for EEG patterns, disruption of information flow, and neurobiological function together. The work also compared to clinical observations during loss of consciousness under general anesthesia.

Zhou {\it et al.} \cite{Zhou2015} used a network model with a single percolation-probability parameter without considering many biological details. 
The model \cite{Zhou2015} consists of a layered fractal expansion of nodes from a single source (input) node to multiple destination (output) nodes. The nodes represent anatomical or functional simple or complex brain units. The directional edges represent structural, functional, or causal communications between nodes. Nodes are arranged into layers in 1-to-4 and 1-to-9 fractalisation schemes. Within each layer, nodes are connected by the small-world topology according to the Watts-Strogatz algorithm \cite{WattsStrogatz1998}. 

The node amplitudes $A_i(t)$ denote neural activity and are given by $A_i(t)=\sum_j w_{ij} P_j(t)$, where the input function $P_j(t) = \sum^m_{ \tau=1} e^{-\tau}A_j(t-\tau)$ represents the history of activity during m previous time steps.
The weights $w_{ij}$ of the directional edges are stochastically defined by a single parameter $p$ representing the percolation probability of information transmission. The system is driven by white noise at the first input node $i=1$, and the output noise is detected at randomly selected single nodes in the topmost layer.
The time-dependent output is analysed in terms Fourier spectra, spectral power and relative spectral entropy \cite{Zhou2015}. 
The network percolation algorithm produces patterns that are similar to human EEG patterns, and broadly describes how the frequency and power distributions of the  $\alpha, \beta, \gamma, \delta$ signals change going from conscious to unconscious states during anesthesia. 
In particular, the model reveals emergence of phase coherence  and synchronisation at a critical threshold.

In summary: Zhou {\it et al.} \cite{Zhou2015} propose that the behaviour of the model supports the notion that consciousness may arise from the same basic statistical processes as those governing the self-organised criticality, regulated by a single connectivity parameter. They state that the results agree with clinical observations of (i) a spectral shift of power toward lower frequency during sedation and (ii) a sharp transition between conscious and unconscious states within an extremely narrow anesthetic concentration range.

One may now wonder what is the relevance of the thalamo-cortical SMN communication network for identifying consciousness?
A way (my way) to see this is to note that the model is generic, including no biological detail. The fractal structure is generic, and should describe the path from any specific "input" node to any other distant "output" node in any of the major networks discussed before: DMN, CEN, SN, CN, SMN. 

Since the model  \cite{Zhou2015} connects the computed EEG time pattern for the model SMN with actual clinical results for the global brain, the result should be relevant for the appearance of phase coherence and  synchronisation in generic large-scale brain networks. Which specific network, if any, is the home of consciousness is a different matter, to be discussed in the next Section.

\newpage

\section {Consciousness and cognition}

\subsection {Current paradigms: GNW and IIT}

For the last two decades,  neurobiological paradigms have been largely dominated by Global Neural Workspace (GNW) theory \cite{Dehaene2017,Baars1988,ChangeuxDehaene1989,Baars2002,Dehaene2001,DehaeneChangeux2011} 
and Integrated Information Theory (IIT) \cite{Tononi2008,Casali2013,Tononi2016,Koch2018,Tegmark2015}. 
Both GNW \cite{Dehaene2017} and IIT \cite{Koch2018} are founded in neurobiology and the functionality of neural circuits, but the way they make use of the common experimental knowledge is fundamentally different. 
Broadly speaking, GNW \cite{Dehaene2017} describes consciousness as a {\it computable} global  property of a neural network, while 
IIT \cite{Koch2018} posits that consciousness is connected with high values of irreducible global network information. 
IIT predicts that a sophisticated simulation of a human brain running on a digital computer cannot be conscious - even if it can speak in a manner indistinguishable from a human being \cite{Koch2018}. This is in stark contrast to the predictions of GNW which posits that machines (AIs) will be able to show consciousness at a number of different high levels \cite{Dehaene2017} independently of the substrate.

Both schools are addressing the same basic experimental facts, and the differences seem rooted in concepts and beliefs developed three decades ago. During the last decades, brain imaging technologies like EEG, MEG, PET and fMRI have made it possible to study the behaviour of the brain in time and space, discovering a wide range of brain patterns connected with all sorts of conditions from vegetative states, via various forms of unconsciousness, via mental diseases, to normal brains and behaviours. Moreover, by stimulating the brain with external magnetic pulses - transcranial magnetic stimulation (TMS) - and recording EEG spectra and/or fMRI images, one can get information about correlations in time and space  \cite{Barttfeld2015,Bestmann2013,Sarasso2014,Hallam2016,Hawco2017,Saari2018,Leitao2017}.

Tononi's information integration theory (IIT) \cite{Tononi2004,Tononi2008} treats consciousness as a global irreducible state of a complex system characterised by a single parameter $\Phi$ describing the information content. The $\Phi$ parameter is effectively obtained by averaging the EEG output from repeated TMS pulses and zip-compressing the output data. Simple regular patterns characterising unconscious states will have low $\Phi$, while conscious states with high activity will show high $\Phi$. 
Koch \cite{Koch2018} refers to it as a "consciousness meter". Maquire {\it et al.} have discussed understanding consciousness as data compression \cite{Maguire2016}, and argued that consciousness may not be computable \cite{Maguire2014}.

IIT is axiomatic and puts itself at the side of Dualism, and there is no bottom-up computational basis for predictions. A critical analysis of IIT is given by Aaronson \cite{Aaronson2014}, with response by Tononi  \cite{Tononi2014}. Further critical analyses of IIT versus Functionalism/GNW have been provided by Cerullo \cite{Cerullo2015}, Fallon \cite{Fallon2016,Fallon2018} and Bayne \cite{Bayne2018}.

In contrast, computational neurobiology is based on concrete models of brain networks, and GNW belongs to this class. Real brain networks have a hierarchical network structure  \cite{Dehaene2017,Barttfeld2015,Meunier2010,Hilger2017}. GNW posits that local structures are able to access collective properties of the entire network, accessing the global workspace, and that this possibly represents consciousness. Like seeing an object (a local visual response) may trigger the recall of a scene, an episode, out of memory.

\subsection{Global dynamics and EEG patterns in models of brain networks}

Arguably, the only systematic, albeit time consuming, way to make progress is to develop and test top-down and bottom-up models against each other, in order to converge to a consistent picture.  The top-down models must correctly describe behavioural patterns. The bottom-up models should describe the corresponding properties of the physical network. Together they must provide verifiable neural correlates of consciousness (NCC) \cite{Fink2016}. The percolation model \cite{Zhou2015} is an example of a top-down global model that captures some essential global properties. 

 It seems that the most rational approach is to define basic consciousness in terms of the real-time dynamics of the default mode DMN resting state network, the highest-level global network, when identified as consciousness in brain imaging.
In this state, the subject is "day-dreaming", thinking about things with all couplings to attention networks effectively turned off. This is at the core of self-awareness, at the stage where psychedelics will dissolve the "self". At the extreme, this could perhaps be a person with completely-locked-in syndrome (CLIS), immersed in thought but unable to communicate (except via the EEG pattern \cite{Murguialday2011,Chaudhary2017,Guger2017,Lesenfants2018,Rohaut2017}). Since the DMN is a global network, there is no reason to associate NCC with any particular module or hub - the proper NCC should be the {\it dynamic global pattern}. Nevertheless, features of the DMN dynamics should be able to read out locally, e.g. at nodes by other networks.

Cognition is quite a different matter, involving {\it attention} and {\it action}. Cognition is defined as  (i) "the mental action or process of acquiring knowledge and understanding thought, experience, and the senses" as well as (ii) "a perception, sensation, idea, or intuition resulting from the process of cognition".
The first definition (i) is operational and tangible, and can be characterised by network activities. The second definition (ii) is abstract:  
the concrete manifestations of subjective experience - qualia - must necessarily be identified via the results of mental {\it action} and subsequent {\it physical information processing}.

\subsection{Default Mode Network - the  key to the self?}

The concept of a top-level Default Mode  Network (DMN; see Sect. 4) was introduced by Reichle et al. in 2001 \cite{Reichle2001} in order  to describe the "introspective mind-wandering" resting state of the brain as evidenced by fMRI and PET  \cite{Reichle2007}. 

As discussed before, there are good reasons for regarding the global DMN as the home of self-consciousness. Ideally it represents the brain and mind of a person totally disconnected from input from the environment. The effects of psychedelics induce a deeper introspection and dissolution of the "self" and can be traced to changes in the connectivity of the DMN \cite{Carhart-Harris2014,Carhart-Harris2017a,Carhart-Harris2017b,Fontes2015}. 

The DMN is often characterised as a "task-negative network" \cite{Reichle2001,Reichle2007} because it "shuts down" when an interrupt signal is requesting attention (like good old microprocessors).
The generic term task-positive network (TPN) is then sometimes used to refer to a collection of attention and action networks (CEN, SN, CN, SMN)  that are anti-correlating with the DMN: when they "turn" on because of some external input, they make the DMN "turn off" ("stop thinking"). The command can of course also come from inside, as a self-conscious decision to pay attention, and even to perform a task.

Cognition implies interaction with the external environment through sensors involving touch (SMN), hearing, seeing and smelling. Cognition also involves interaction with the internal environment, e.g. with memory,  through pure contemplation. An external sound or an internal bright idea will excite the CEN to demand attention and action via the DMN. The internal environment involves the salience network (SN) and the cerebellum (CN) expressing perception, emotions, fear, anger.
Evolution must have used it to reward us and to save us from danger \cite{Wagner2017,Hausknecht2017,Clausi2017,Fahrenfort2017,Roy2017,Thompson2018}. 

To be able to build a network that displays some elementary form of consciousness that can be identified in operational terms (e.g. EEG pattern), it then seems reasonable to expect that one "only" needs a minimal model of the brain containing the essential collective (global) and local network properties. This may have nothing to do with human consciousness, but that is a different matter. To go on, one can include elementary versions of major components of  the TPN to give the system memory and some ability of self-monitoring and self-reflection. It may be a formidable task, but it seems doable.

\subsection{Levels and disorders of consciousness}

"Level of consciousness (LoC)" is an exceedingly complex concept. It is at the core of much of the philosophical confusion, but also, and more importantly, is at the focus of current clinical medical attention \cite{Sergent2017,Naccache2018}. "Disorders of Consciousness (DoC)" range from coma and vegetative state (VS)/unresponsive wakefulness (UWS) through minimally conscious states (MCS) to full awareness and cognition. This is then to be compared with normal states of sleep and states induced by sedation and drugs (including psychedelics). The MCS-label is problematic, because MCS is not a well-defined state: it covers a broad range of responses and needs to be "expanded" in a multi-dimensional space of brain patterns and behaviours \cite{Sergent2017,Naccache2018,Bayne2016,Bayne2017}.

In a recent review in Science: "What is consciousness, and could machines have it?", Dehaene, Lau, and Kouider \cite{Dehaene2017} discuss how to define different LoCs, and how artificial systems might perceive things. 
In particular Dehaene {\it et al.} \cite{Dehaene2017} propose that the word "consciousness" combines two different types of information-processing computations in the brain: the selection of information for global broadcasting, thus making it flexibly available for computation and reporting (C1), and the self-monitoring of those computations, leading to a subjective sense of certainty or error (C2). Moreover, Dehaene et al. argue, not surprisingly, that current machines are still mostly implementing computations that reflect unconscious processing (C0) in the human brain.

A recent example could be AlphaGo Zero, the AI that is the new master of the game of Go  \cite{Silver2017}.
A way to characterise the ability of machines might be to apply the emerging multi-dimensional knowledge and taxonomy of "Disorders of Consciousness (DoC)" \cite{Sergent2017,Naccache2018,Bayne2016,Bayne2017} to descibe their behaviour. Like "What can a human do that a machine can't"?  Actually, one might argue that most challenging question might be "What are intentions, emotions, and common sense, and could machines have them?"

For AlphaGo Zero to reach the C1 level, it must be able to explain how it arrived at various decisions during a game of Go and report to the world - this needs episodic memory. And to reach level C2 it must be able to remember its reasoning and critically analyse its behaviour - this takes a lot of memory and higher network functions. This memory and higher functions is the crux of the matter, still to be developed and evolved.

 \subsection{From coma to completely-locked-in syndrome}

Seen from the outside, coma/VS/UWS and completely-locked-in syndrome (CLIS) look similar - no signs of motor response from the patient to any external input. On the inside, however, VS/UWS \cite{Espejo2015} looks very different from CLIS \cite{Murguialday2011,Chaudhary2017,Guger2017,Lesenfants2018,Rohaut2017}.

A completely-locked-in state (CLIS) is typical for the final state of ALS, but can also result from stroke. CLIS is characterised by broken connections with the output circuits, specifically the motor circuits - the person is completely paralysed. ALS is a complex case of neurodegeneracy \cite{VanDamme2017}, with motor neurons and axons dying in the motor cortex and/or in the spine.  After stroke, the  muscles and nerves are still there, but important motor regions of the brain are damaged.
Nevertheless, CLIS patients can hear, which means they can show attention and  respond internally "in the mind"  to commands, but not activate the SMN.

If we take the experimentally established small-world-network view, the brain consists of a number of tightly connected local nodes coupled together to a global network. Moreover, many of these local nodes are connected to internal and external sensors for monitoring the environment. This means that they together could produce a number of collective (global) signals, e.g. a variety if EEG patterns, including neural correlates for perceptual and attentional states  \cite{List2017}.
It also means that perception and awareness may critically depend on the sensor environment - multisensory perception \cite{Deroy2016,Deroy2014,OCallaghan2017,Briscoe2017}.

\subsection {Reading the brain with image processing and machine learning}

Neural Correlates of Consciousness (NCC) necessarily involve time dependent brain patterns, resulting from intrinsic or extrinsic actions and detected with EEG, MEG, PET, fMRI, fNIRS \cite{Chaudhary2017}, etc.  Structural patterns will underlie it all, but dynamic patterns are essential for NCC. 
This is particularly important because it opens the way for direct communication with persons with severe disabilities \cite{Chaudhary2017,Guger2017,Lesenfants2018,Espejo2015,Oken2014}. 

To make use of the time-dependent EEG spectrum one needs to develop useful measures to extract the most relevant information. A commonly used measure is spectral entropy in various forms \cite{Zhou2015,Lesenfants2018,Liang2015,Wirsich2018,Vidaurre2018,Oken2014,Tian2017,Mateos2017}. However, one can anticipate that advanced methods to extract time-dependent correlations will eventually be developed. 

Important progress has been made possible by the recent development in real-time spectral and image processing using machine learning (ML) based on deep neural networks \cite{Schirrmeister2017,Wen2018,Hong2018}.  Recent examples involve cognitive functioning in fragile-X syndrome \cite{Knoth2018},  Alzheimer's disease \cite{Simpraga2017}, detection of depression \cite{Cukic2018}, and predicting sex from brain rhythms  \cite{vanPutten2018}.

Recently Shen {\it et al.} \cite{Shen2017,Shen2018} performed deep image reconstruction by recording human brain activity. A picture of an animal was shown to a person and to a camera. A deep ANN was trained to identify the pixelated camera picture and compare with the fMRI signal pattern. In the end, the trained ANN was able to display, pixel by pixel, unknown pictures shown to the human. The accuracy was low, but there was some clear resemblance. 

In related work, Fong {\it et al.}  \cite{Fong2018} used human brain activity to guide a machine learning algorithm,
taking fMRI measurements of human brain activity from subjects viewing images, and infusing the data into the training process of an object recognition learning algorithm. After training, the neurally-weighted classifiers were able to classify images without requiring any additional neural data \cite{Fong2018}.

It follows that EEG/fMRI readout plus real-time data analysis and image processing plus deep ML with ANNs may provide amazing opportunities for feedback and control in both directions: the real brain can train the ANN, and the ANN can train the brain. Training a human brain could take the form of interaction via the normal senses, as well as through artificial sensors, including implanted ones. An example is an implanted electrode array picking up signals from the motor cortex of an ALS patient \cite{Vansteensel2016}, replacing part of the brain's sensorimotor system by an artificial one.

\newpage

\section{Quantum cognition }

Cognition is the (human) mental process of acquiring knowledge and understanding via thought, experience, and senses.
If consciousness and cognition are global emergent phenomena in a sufficiently complex biological brain, how can cognition be influenced by quantum physics? And, would it make any difference, would we even notice? 

If one searches for "Quantum consciousness" one gets right into a world of quantum mysticism, quantum mind and quantum healing. If one instead searches for  "Quantum cognition" one meets an active research field in psychology  \cite{Wang2013,Lukasik2017,Broekaert2017,Pothos2017,BarrosOas2017},  applying quantum probability theory to human behaviour to model the phenomenology of human decision making:  contextuality (experimental conditions), uncertainty, incompatibility, and dependence on order. An alternative view \cite{BarrosOas2017} suggests that contextual effects may be better modelled by allowing non-observable probabilities to take negative values to describe certain contextual incompatible stochastic processes, rather than describing them by quantum probabilities. 

Further search along the lines of "Quantum games",  "Quantum decisions" and  "Quantum finance" opens up a world of quantum applications for rational decision making \cite{Flitney2002,BrunnerLinden2013,Bang2016,BusemeyerBruza2012,White2016,Martinez2016,Baaquie2013}.
All these fields are variants of optimisation problems, usually involving Ising-type quantum spin Hamiltonians generalising classical binary problems. To provide quantum advantage (i.e. exponential speedup) the algorithms must run on quantum hardware.

\subsection{Quantum effects in the brain - would we even notice?}

But what about quantum enhanced cognition in biological brains?
Arguably, from a materialist/functionalist point of the potential power of quantum cognition must be related to a corresponding classical biological model of cognition. One needs to start from a reasonable model for consciousness and cognition and argue for how physical quantum effects will improve performance. 

There is a variety of ideas how quantum effects might influence or control perception, consciousness and cognition (see e.g. \cite{Jedlicka2017,Bourget2004,Georgiev2015,Georgiev2018} for discussion and references). In some scenarios, quantum effects in the brain tend to be linked to local, or even global, quantum coherent processing and entanglement, but few of them are anchored in realistic brain models of biology and cognition.  

As discussed in Sect. 5, experimentally based  realistic network models of the brain have been developed during the last two-three decades. However, issues of neural correlates of consciousness (NCC) have tended to depreciate those models as reductionist, unable to explain the essential hard questions. 

This has now arguably changed for good: the development during the last three to five years has confirmed basic network models, demonstrated global dynamic brain-like behaviour and identified reasonable NCCs. We therefore now have concepts and  tools -  levels of consciousness (LoC) and disorders of consciousness (DoC) (Sect. 5) - for discussing and assessing in a verifiable manner how quantum effects might influence - but not create! - consciousness and cognition.   

\subsection{Biochemical quantum effects on cognition}

At the local biochemical molecular level, it is all quantum. Quantum tunnelling,  interference and entanglement may play decisive roles in biochemical reactions via  intermediate transition-state complexes. Quantum effects underlie protein folding, and the functioning of enzymes and ribosomes. Malfunction and destruction of neurons and nerves in ALS \cite{VanDamme2017} might be influenced by biochemical quantum effects.  Nevertheless, these are "trivial quantum effects" in the context of consciousness and cognition.

On the contrary, the balance between two types of serotonine receptors in the cortex \cite{Carhart-Harris2017b} is perhaps the best example of how biochemical quantum effects can directly influence consciousness and cognition ("dissolution of the self") by modifying the functional connectivity in the default mode network (DMN). This can then be generalised to all sorts of chemical agents and external probes that modify the functional connectivity of the DMN, globally or locally, perhaps also via coupling to and feedback from the task-positive attention networks (TPN).

The nature and range of coherence in time and space in receptor complexes is still an open question (see Jedlicka \cite{Jedlicka2017} for a recent review of quantum biology). 
However, ideas of {\it macroscopic} quantum coherence in brain networks remain pure speculation and are completely unwarranted.

\subsection{Fisher's idea of flying spin qubits}

Nevertheless, a recent paper by Fisher \cite{Fisher2017} presents an interesting proposal for how it might be possible to influence chemistry in the brain by entangled quantum processes involving well-protected  flying spin qubits. Fisher's approach can be viewed as a bottom-up approach, developing a concrete mechanism for coherent quantum processing based on physical long-coherence quantum components, with no need for quantum error correction. The model explicitly deals with quantum computing in the brain, encapsulating phosphorous spin qubits in phosphate ions, paired into "Posner molecules" to provide potentially long-lived spin qubits, entangled pairs, and quantum memory (many seconds).

Fisher argues that the enzyme-catalysed chemical reactions can entangle spin qubit pairs, and that the pairs can be transported to synapses and released in such ways that resulting neurotransmitter cascades are entangled and give rise to non-local correlations. Fisher takes the capability of modulating the excitability and signalling of neurons as a working definition of "quantum cognition".

The obvious objection is the same as before:  the proposed coherent processes can possibly locally entangle some components or nodes, but hardly globally entangle the brain network. It might influence rate coefficients in the classical brain network, but hardly provide anything like quantum computing, or genuine "quantum cognition" that would not exist in the classical brain.

Even if Fisher's quantum processes do not exist in the brain, it seems technically possible to build artificial chemical systems that might display such effects - that the outcome of the chemical reactions depend on entanglement created at some point. 
This is similar to the proposed mechanism for magnetoreception \cite{CaiPlenio2013,Tiersch2014,Mohseni2014}.

\subsection{Stapp's view of a quantum brain}

Stapp takes a top-down dualist view, arguing for a global quantum brain where coherence is maintained by mental action \cite{Schwartz2005,Stapp2011,Stapp2017}. 
Stapp's core argument is that basic brain components are best understood in quantum terms, and the brain itself must be treated as a quantum system. This evidently leads to the need for a quantum coherent brain, and Stapp developed a theory describing the effects of mental action on brain activity as achieved by a Quantum Zeno Effect that is not weakened by decoherence (see \cite{Georgiev2015} for an opposing view). Thus, terms having intrinsic mentalistic and/or experiential content (qualia), like "feeling", "knowing" and "effort", are included as primary causal factors. Nevertheless, like in all other theories, there is really no functional relation between the quantum brain network and qualia, modelling the emergence of feelings in a quantum network.

\subsection{Penrose/Hameroff and  orchestrated objective reduction}

During the 1980s, Stuart Hameroff proposed a classical version of protein-connected microtubules as classical molecular automata performing basic computational logic of the living state \cite{Hameroff1988} . In that work, there was no mention of quantum oscillators, and it was in line with other ideas within the fields of molecular computing and artificial life. However, the collaboration with Penrose took microtubules to a hypothetical quantum level. In his 1994 book "Shadows of the Mind", Penrose \cite{Penrose1994} hypothesised that human consciousness is non-algorithmic (non-computable) and therefore cannot be simulated by a universal Turing machine, let alone a classical digital computer. 

Penrose chose to connect his view of non-computable consciousness with his views on the foundation of quantum mechanics: objective reduction (OR) postulates that quantum gravitational interaction at the Planck scale will inevitably make quantum wavefunctions collapse physically, and that the collapse is non-computable (but still physical!). Penrose apparently regarded quantum mechanics to be fundamentally non-computable, thus providing a tool for understanding a likewise non-computable brain.

In a subsequent 1996 paper \cite{HameroffPenrose1996} Hameroff and Penrose introduced microtubule networks as quantum coherent networks performing quantum computing. They postulated that resultant self-collapse of the quantum network wavefunction (orchestrated objective reduction, Orch-OR), irreversible in time, creates an instantaneous "now" event, and that sequences of such events create a flow of time, and consciousness \cite{HameroffPenrose1996}. It is interesting to note that the Orch-OR picture does not require very long coherence times - "only" in the 100 millisecond range corresponding to the time frame for classical nervous processes and human cognition. However, in 2000 Tegmark \cite{Tegmark2000} calculated that the coherence time in microtubules would fall in the picosecond range or shorter (requiring a very fast-thinking brain). 

Even if a "warm, wet and noisy" environment can allow longer coherence times, milliseconds is still a very long time. Moreover, it does not really matter: it all hinges on Penrose postulating that non-computational quantum mechanical ingredients are present in nature, and that the wavefunction collapse - not the quantum computation itself - causes the quality of consciousness. If this were true, which is not likely, there would necessarily exist some physical mechanism to be found. 

Most other models are based on efficient quantum computation in a computable brain, and then very long coherence times become essential. 

\subsection{Back to reality - the classical brain}

At this point, the still engaged reader may ask why bother arguing about theories and hypotheses that one anyway rejects? Views that can be regarded as outdated and/or manifestly unrealistic? The answer is that many of these ideas about quantum physics, human mind and free will are still highly visible, and often promoted by highly respected, famous and influential scientists and philosophers. They present topics for deep and  interesting discussions, and the popular books describing the ideas are often highly rated. 

The problem is that they also contribute to creating an artificial conflict between up-to-date materialism/functionalism and quantum physics. It is time to develop narratives that are driven by a genuine interest in presenting modern views on how to integrate functionalism and quantum physics with consciousness and cognition, rather than promoting far-out views based on often speculative ways of how to interpret quantum mechanics. 
The discussion in the first part of this section hopefully provides a realistic view of how local quantum effects on the functional connectivity of classical brain networks may influence consciousness and cognition.

\newpage

\section {Quantum neural networks - quantum brains ?}

The brain is certainly not a quantum computer, and significant quantum-enhanced functionality influencing consciousness and cognition in biological brains is very unlikely. However, small quantum computers and simulators are emerging: it is now possible to implement quantum algorithms and protocols on systems with 10-20 qubits \cite{Wendin2017}, and operational systems with 50-100 qubits will be available in the immediate future \cite{Wendin2017,Preskill2018}. Such systems form real-space physical qubit networks, and the gate operations creates networks in state (Hilbert) space. 

The notion of "Quantum brain" refers to quantum neural networks (QNN) and recurrent QNNs (RQNN). It may be anything from a simple network of connected qubits to networks of important complexity, e.g. a hierachial distributed network of quantum nodes that are networks in themselves. The obvious question is: can such quantum networks mimic a biological brain, combining the strengths of neural processing with the exponentially larger state space and potential algorithmic speedup of quantum systems to achieve some form of quantum AI - QAI? 
A generic answer is "yes, in principle" - and QNNs are obvious platforms for hybrid analogue-digital QC (ADQ). 

Actually, the present and near-future NISQ processors \cite{Preskill2018} face similar problems as biological brains: in practice a "noisy, warm and wet" environment. A major problem for quantum computing is decoherence and loss of entanglement. This means that the present applications of digital algorithms necessarily can only have low depth and a rather small  number of gates.  

In classical CMOS hardware is now possible to physically configure chips to function as a network of neurons and synapses \cite{Merolla2014,Maass2016a}. One can also design and simulate the neural network and all of its functions in software (see e.g. Human Brain project - HBP)  \cite{HBP2018}. 
In principle one could imagine a similar quantum software approach if one has access to a large coherent QC, e.g. simulating quantum perceptrons  \cite{Schuld2015,Wiebe2016}, and synaptic weights and neurons \cite{Cao2017,Wan2017}. However, this may be neither realistic nor efficient: such large QCs will need error correction, and are for the distant future, if at all. A digital QC approach may not be particularly useful for brain-like applications. There are however interesting proposals for building QNN components in hardware \cite{Pfeiffer2016,Salmilehto2017,Cheng2018,Silva2018}.

\subsection {Quantum reservoir computing (QRC)}

In Sect. 4 we described reservoir computing (RC) with recurrent classical neural networks as a very promising approach to modelling the dynamics and functionality of components of a brain. As discussed by Enel {\it et al.} \cite {Enel2016}, mixed selectivity in cortical neurons \cite{Fusi2016}, representing a variety of situations defined by current sensor inputs, is readily obtained in randomly connected RNNs. In this context, these reservoir networks reproduce the highly recurrent nature of local cortical connectivity \cite {Enel2016}.

It is then natural to investigate the computational power of complex dynamical quantum systems in a similar  way using recurrent quantum networks - RQNN.
Quantum reservoir computing (QRC) might be particularly useful for quantum information processing because loss of memory - fading memory - is a key RC resource, and there is no need for global control of the reservoir network. 
Present and emerging quantum hardware with limited coherence should be able to serve as a fading-memory reservoir for QRC. 

The field is just opening up, and there are so far only a few papers on the topic.
Fujii and Nakajima \cite{Fujii2017} have performed numerical QRC experiments that show that quantum systems consisting of 5-7 qubits possess computational capabilities comparable to conventional RNNs with 100-500 nodes.
Nakajima {\it et al.}  \cite{Nakajima2018}  introduce a spatial multiplexing scheme to boost the computational power via multiple different quantum systems driven by common input streams in parallel.
Kutvonen {\it et al.}  \cite{Kutvonen2018}  study memory capacity and accuracy of a QRC based on the fully connected transverse field Ising model. They show that variation in inter-spin interactions leads to a better memory capacity in general, by engineering the type of interactions the capacity can be greatly enhanced and there exists an optimal timescale at which the capacity is maximised.

\subsection {Machine learning in feed-forward and recurrent QNNs}

There is a huge number of recent papers using machine learning (ML) to train artificial neural networks to implement algorithms and protocols, e.g. for classifying data. There is an excellent review article by Biamonte {\it et al.} \cite{Biamonte2017} which, together with a number of recent research papers on Hopfield recurrent QNNs \cite{Rebentrost2017,Rotondo2018}, quantum Boltzmann machines \cite{Amin2016,Deng2017}, quantum circuit QNNs \cite{Mitarai2018,FarhiNeven2018}, and quantum agents and artificial intelligence \cite{Briegel2012,Mautner2015,Dunjko2015,Dunjko2016,Melnikov2017,Melnikov2018}, covers most of the field. 

Here we will only briefly discuss two papers \cite{Mitarai2018,FarhiNeven2018} that promise applications for "Noisy Intermediate-Scale Quantum" (NISQ) processors \cite{Preskill2018} by generating QNNs as sequences of quantum circuit gates implemented in a "standard" quantum computer. 

Farhi and Neven \cite{FarhiNeven2018} construct the quantum circuit as a sequence of parameter dependent unitary transformations which acts on an input quantum state. 
For classification of binary data, the input quantum state is an n-bit computational basis state corresponding to a sample string. 
Farhi and Neven \cite{FarhiNeven2018} show how to design a circuit made from two qubit unitaries that can correctly represent the label (effectively parity) of any Boolean function of n bits. Through classical simulation the parameters can be found that allow the QNN to learn to correctly distinguish the two data sets.  Farhi and Neven \cite{FarhiNeven2018} anticipate that it will be possible to run this
QNN on a near term gate model quantum computer where its power can be explored beyond what can be explored with simulation. 

Mitarai {\it et al.} \cite{Mitarai2018} propose a hybrid classical-quantum algorithm for machine learning - quantum circuit learning (QCL) -  that they suggest can provide an alternative to the Variational Quantum Eigensolver  (VQE) and the Quantum Approximate Optimization Algorithm  (QAOA). 
QCL  and QRC both pass the central optimisation procedure to a classical computer. 
In QRC, the output is defined by a set of observables taken from quantum many-body dynamics driven with a fixed Hamiltonian, and the weight vector is tuned on a classical device to minimise the cost function. 
In contrast, QCL  tunes the whole network and can be regarded as a QNN with tunable couplings, the quantum counterpart of a basic neural network.

\newpage

\section{Perspectives}

Thomas Nagel's book "Why the Materialist Neo-Darwinian Conception of Nature is Almost Certainly False" \cite{Nagel2012} played a pivotal role in forming my view of how philosophers underestimate the power of science, especially classical physics. The view in that book is that even the modern materialist approach to life fails to explain consciousness, intentionality, meaning, and value. Therefore, according to Nagel, this threatens the entire naturalistic world picture, extending to biology, evolutionary theory, and cosmology. The simple fact is, of course, that we have no idea of whether modern materialist approaches to life fail before they provably fail. And that may take a long time and lots of patience to find out. 

Or maybe not: the recent experimental and modelling progress to understand the brain's global and local network structure and dynamics is amazing: the "hard problem of consciousness" has dissolved, and the classical complexity of the brain is huge but not overwhelming. There is no reason to consider quantum complexity: quantum effects are basically biochemical, able to influence both the local and global  functional connectivity of brain networks. This can evidently influence consciousness, cognition and motor responses, but has nothing to do with a "quantum mind". An individual's subjective experience of "value, intention and meaning" is the result of internal classical information processing of interacting networks, comparing the various physical experiences of that individual. Functional network disorder can then obviously create highly personal views of  "value, intention and meaning".

It is interesting to note that the most visible actors have often been people in quantum physics of particles and fields who have a strong interest in the "physics" of the mind, or people in biology, medicine and philosophy with a strong interest in esoteric concepts in quantum physics, like superposition, coherence, entanglement, measurement and wavefunction collapse. As a result, views of quantum physics have often bordered to quantum mysticism, and been used to support panpsychism and a universal mind based on tentative quantum processes in biological brain components. 

The optimistic conclusion of this paper is that we may soon be facing a paradigm shift where some basic mechanisms of consciousness and cognition can be efficiently described by classical analogue and digital systems. The problem will then be to understand the spectrum of consciousness and self-awareness, from bacteria and {\it C. elegans} via Lucy and modern man, to future AIs that report experiences and emotions. 

At the present level of knowledge in neurobiology, quantum processes probably play no significant role for cognition. However, this can change in future artificial cognitive systems involving quantum circuits, with quantum agents living in quantum networks and demonstrating free will \cite{Briegel2012}. Nevertheless, even these powerful quantum agents \cite{Briegel2012} will still have limited power - Nature will only solve really hard problems approximately, and quantum computers may not always have any advantage.

There is an old Swedish proverb: "När fan blir gammal blir han religiös" ("When the devil grows old he becomes religious"). Closing in, we tend to challenge established views,  sometimes moving to the extreme edges, searching for the meaning of it all. The radical new insight is that we might in fact be able to understand who we are, given a bit of time. 

For some people, the year is 2023. In a recent article in New Scientist a Swedish science writer describes an interesting encounter in Europe 1998.  Quoting Per Snaprud  \cite{Snaprud2018}: {\it "Twenty years ago this week, two young men sat in a smoky bar in Bremen, northern Germany. Neuroscientist Christof Koch and philosopher David Chalmers had spent the day lecturing at a conference about consciousness, and they still had more to say. After a few drinks, Koch suggested a wager. He bet a case of fine wine that within the next 25 years someone would discover a specific signature of consciousness in the brain. Chalmers said it wouldn't happen, and bet against."}

The core of the bet is to demonstrate a convincing neural correlate of consciousness (NCC) by 2023. For Koch it would be an experimental "consciousness meter", providing convincing response seen in EEG brain waves after zapping the brain with an electromagnetic pulse \cite{Koch2017,Koch2018}.

The recent development, as described in Sect. 5, suggests that David Chalmers will have to buy the wine. However, they should drink the wine together, and invite Stanislas Dehaene to join, because Christof Koch will not really win. The delocalised global default mode network (DMN) is verifiably the best model for the home of consciousness, and the NCC is given by the dynamics of the DMN resting state as manifested in the information content (entropy) of the EEG spectrum. In practice, the NCC is already used clinically to map disorders of consciousness and to create methods for two-way communication, training and control. This will probably lead to a quite dramatic development during the next decades, assisted by a parallel development of machine learning and AI. However, one can expect  exciting progress already in the near future - there will be good reasons for Koch and Chalmers to share a case of wine in 2023.

\newpage

\section*{Acknowledgement}

This work has been partially supported by the Knut \& Alice Wallenberg Foundation (KAW)  (WACQT project), and by the EU Quantum Technologies Flagship (820363 - QpenSuperQ).


\newpage

\section *{References}


\begin{thebibliography}{99}

\bibitem{Aaronson2005}
S. Aaronson,  NP-complete problems and physical reality?, {\it ACM SIGACT News} {\bf2005}, {\it36}, 30.

\bibitem{Aaronson2013}
S. Aaronson, Why philosophers should care about computational complexity,  in {\it Computability: G\"odel, Turing, Church, and Beyond} (Eds. B. Copeland, C. J. Posy, O. Shagrir), MIT Press, Cambridge, MA {\bf2013}, pp. 261-328.

\bibitem{Broersma2018} 
H. Broersma, S. Stepney, G. Wendin, Computability and Complexity of Unconventional Computing Devices, in {\it Computational Matter}, (Eds. S. Stepney, S. Rasmussen, M. Amos), Springer {\bf2018}; arXiv:1702.02980v2.

\bibitem{Wendin2017} 
G. Wendin, Quantum information processing with Superconducting circuits: a review, {\it Rep. Prog. Phys.} {\bf2017},  {\it 80}, 106001; arXiv: 1610.02208v2.

\bibitem{Chalmers1995}
D. J. Chalmers, Facing up to the problem of consciousness, {\it J. Consciousness Studie}s {\bf2017}, {\it2}, 200.

\bibitem{Bak1996}
P. Bak,  {\it How Nature Works: The Science of Self-Organised Criticality,} New York, NY: Copernicus Press {\bf1996}.

\bibitem{Bullmore2009}
E. Bullmore, A. Barnes, D. S. Bassett, A. Fornito, M. Kitzbichler, D. Meunier, J. Suckling,
Generic aspects of complexity in brain imaging data and other biological systems,
 {\it NeuroImage}  {\bf2009},  {\it47}, 1125.

\bibitem{Bassett2011}
D. S. Bassett, M. S. Gazzaniga, 
Understanding complexity in the human brain,
{\it Trends in Cognitive Sciences} {\bf2009}, {\it15}, 200.

\bibitem{Chialvo2010}
D. R. Chialvo, Emergent complex neural dynamics, {\it Nature Phys}  {\bf2010}, {\it 6}, 744.

\bibitem{Tagliazucchi2016}
E. Tagliazucchi, D. R. Chialvo, M. Siniatchkin, E. Amico, J.-F. Brichant, V. Bonhomme, Q. Noirhomme, H. Laufs, S. Laureys, Large-scale signatures of unconsciousness are consistent with a departure from critical dynamics, {\it J. R. Soc. Interface} {\bf2016}, {\it13}, 20151027.

\bibitem{Tagliazucchi2017}
E. Tagliazucchi, The signatures of conscious access and its phenomenology are consistent with large-scale brain communication at criticality, 
{\it Consciousness and Cognition} {\bf2017}, {\it55}, 136.

\bibitem{Lake2017}
B. M. Lake, T. D. Ullman, J. B. Tenenbaum, S. J. Gershman, Building machines that learn and think like people, {\it Behavioral and Brain Sciences} {\bf2017}, {\it40}, e253.

\bibitem{Cocchi2017}
L. Cocchi, L. L. Gollo, A. Zalesky, M. Breakspear, Criticality in the brain: A synthesis of neurobiology, models and cognition, 
 {\it Progress in Neurobiology}  {\bf2017},  {\it158}, 132. 

\bibitem{Maass2016}
W. Maass, Searching for principles of brain computation, 
{\it Current Opinion in Behavioral Sciences} {\bf2016}, {\it11}, 81.

\bibitem{ChangeuxDehaene1989}
J.-P. Changeux, S. Dehaene, Neuronal models of cognitive functions,
{\it Cognition} {\bf1989}, {\it33}, 63.

\bibitem{Dehaene2001}
S. Dehaene, L. Naccache, Toward a cognitive neuroscience of consciousness: basic evidence and a workspace framework, {\it Cognition} {\bf2001}, {\it79}, 1.

\bibitem{DehaeneChangeux2011}
S. Dehaene, J.-P. Changeux,
Experimental and Theoretical Approaches to Conscious Processing,
{\it Neuron}, {\bf2011}, {\it70}, 200.

\bibitem{Dehaene2014}
S. Dehaene, {\it Consciousness and the Brain: Deciphering How the Brain Codes Our Thoughts,}
{\bf2014} Penguin Publishing Group.

\bibitem{Dehaene2017}
S. Dehaene, H. Lau, S. Kouider,
What is consciousness, and could machines have it?,
{\it Science} {\bf2017}, {\it358}, 486.

\bibitem{Barttfeld2015}
P. Barttfeld, L. Uhrig, J. D. Sitt, M. Sigman, B. Jarraya, S. Dehaene,
Signature of consciousness in the dynamics of resting-state brain activity,
{\it PNAS}, {\bf2015}, {\it112}, 887.

\bibitem{KochHepp2006}
C. Koch, K. Hepp, Quantum mechanics in the brain,
{\it Nature} {\bf2006}, {\it440}, 611.

\bibitem{Tononi2004}
G. Tononi, An information integration theory of consciousness,
{\it BMC Neurosci.} {\bf2004}, {\it5}, 42. 

\bibitem{Tononi2008}
G. Tononi, Consciousness as integrated information: a provisional manifesto, 
{\it Biol. Bull.} {\bf2008}, {\it215}, 216.

\bibitem{Casali2013}
A. G. Casali, O. Gosseries, M. Rosanova, M. Boly, S. Sarasso, K. R. Casali, S. Casarotto, M.-A. Bruno, S. Laureys, G. Tononi,  M. Massimini,, A theoretically based index of consciousness independent of sensory processing and behavior, 
 {\it Sci. Transl. Med.}  {\bf2013,}  {\it5}, 198ra105.

\bibitem{Tononi2016}
G. Tononi, M. Boly, M. Massimini, C. Koch, Integrated information theory:  from consciousness to its physical  substrate, {\it Nature Reviews: Neuroscience} {\bf2016}, {\it17}, 450.

\bibitem{Koch2017}
C. Koch, How to make a consciousness meter, Sci. Am., November 2017.

\bibitem{Koch2018}
C. Koch, What Is Consciousness?,  {\it Nature} {\bf2018}, {\it557}, S9.

\bibitem{Dennett1991}
D. C. Dennett,   {\it Consciousness Explained}, Little, Brown and Co. {\bf1991}.

\bibitem{Dennett2017}
D. C. Dennett,  {\it From Bacteria to Bach and Back: The Evolution of Minds}, W. W. Norton \& Company, New York/London {\bf2017}.

\bibitem{Dennett2018}
D. C. Dennett, Facing up to the hard question of consciousness,
{\it Phil. Trans. R. Soc. B} {\bf2018}, {\it37}, 20170342.

\bibitem{Nagel1974}
T. Nagel, What is it like to be a bat?,
{\it Philos. Rev.} {\bf1974}, {\it83}, 435.

\bibitem{Nagel2012}
T. Nagel,  {\it Mind and Cosmos: Why the Materialist Neo-Darwinian Conception of Nature is Almost Certainly False}, Oxford University Press {\bf2012}.

\bibitem{Nagel2017}
T. Nagel, Is Consciousness an Illusion?, The New York Review of Books, March 9, {\bf2017}.

\bibitem{CompMatter2018}
{\it Computational Matter},
(Eds. S. Stepney, S. Rasmussen, M. Amos), Springer {\bf2018}.

\bibitem{Adleman1994}
L. M. Adleman, Molecular computation of solutions to combinatorial problems, {\it Science} {\bf1994}, {\it266}, 1021.

\bibitem{Siegelmann1995}
H. Siegelmann, Computation beyond the Turing limit,  {\it Science} {\bf1995}, {\it268}, 545.

\bibitem{SiegelmannSonntag1995}
H. T. Siegelmann, E. Sonntag, On the Computational Power of Neural Nets, 
 {\it J. Computer and System Sciences} {\bf1995}, {\it50}, 132.

\bibitem{Cabessa2011}
J. Cabessa and H. T. Siegelmann, Evolving recurrent neural networks are super-Turing?, {\it Proc. IJCNN 2011, IEEE} {\bf2011}, 3200.

\bibitem{DiVentraTraversa2018}
M. Di Ventra, F. L. Traversa, Perspective: Memcomputing: Leveraging memory and physics to compute efficiently, 
{\it J. Appl. Phys.}  {\bf2018}, {\it123}, 180901.

\bibitem{Pei2017}
Y. R. Pei, F. L. Traversa, M. Di Ventra,
On the Universality of Memcomputing Machines, 
 {\it IEEE Transactions on Neural Networks and Learning Systems} {\bf2018}; DOI: 10.1109/TNNLS.2018.2872676.

\bibitem{Manukian2018}
H. Manukian, F. L. Traversa,  M. Di Ventra, Accelerating Deep Learning with Memcomputing, {\it Neural Networks} {\bf2019} {\it110}, 1.

\bibitem{Sheldon2018}
F. Sheldon, P. Cicotti, F. L. Traversa,  M. Di Ventra,
Stress-testing memcomputing on hard combinatorial optimization problems (2018); arXiv:1807.00107v1

\bibitem{Traversa2018}
F. L. Traversa, P. Cicotti, F. Sheldon,  M. Di Ventra,
Evidence of Exponential Speed-Up in the Solution of Hard Optimization Problems,
{\it Complexity} {\bf2018}, {\it 2018}, 7982851

\bibitem{Sussillo2014}
D. Sussillo,
Neural circuits as computational dynamical systems,
 {\it  Current Opinion in Neurobiology} {\bf2014}, {\it25}, 156.

\bibitem{Carmantini2015}
G. S. Carmantini, P. beim Graben, M. Desroches and S. Rodrigues,
Turing Computation with Recurrent Artificial Neural Networks,
{\it CoCoNIPS}, {\bf2015}; arXiv:1511.01427.

\bibitem{Vergis1986}
A. Vergis, K. Steiglitz,  B. Dickinson, The complexity of analog computation, Mathematics and Computers in Simulation, {\it Mathematics and Computers in Simulation} {\bf1986}, {\it28}, 91.

\bibitem{Denef2007}
F. Denef, M. R. Douglas, Computational complexity of the landscape: Part I,  
{\it Annals of Physics} {\bf2007}, {\it322}, 1096.

\bibitem{Douglas2013}
K. Douglas, Learning to Hypercompute? An Analysis of Siegelmann Networks,  {\it in Computing Nature: Turing Centenary Perspective} (Eds. G. Dogic-Crnkovic, R. Giovagnoli), Springer {\bf2013}, pp. 201-211.

\bibitem{Merolla2014}
P. A. Merolla, J. V. Arthur, R. Alvarez-Icaza, A. S. Cassidy, J. Sawada, F. Akopyan, B. L. Jackson, N. Imam, C. Guo, Y. Nakamura, B. Brezzo, I. Vo, S. K. Esser, R. Appuswamy, B. Taba, A. Amir, M. D. Flickner, W. P. Risk, R. Manohar, D. S. Modha,
A million spiking-neuron integrated circuit with a scalable communication network and interface,
{\it Science} {\bf2014}, {\it345}, 668.

\bibitem{Esser2016}
S. K. Esser, P. A. Merolla, J. V. Arthur, A. S. Cassidy, R. Appuswamy, A. Andreopoulos, D. J. Berg, J. L. McKinstry, T. Melano, D. R. Barch, C. di Nolfo, P. Datta, A. Amir, B. Taba, M. D. Flickner, D. S. Modha,
Convolutional networks for fast, energy-efficient neuromorphic computing,
{\it PNAS} {\bf2016}, 1{\it13}, 11441.

\bibitem{Maass2016a}
W. Maass, Energy-efficient neural network chips approach human recognition capabilities, {\it PNAS} {\bf2016}, {\it113}, 11387.

\bibitem{Amir2017}
A. Amir, B. Taba, D. Berg, T. Melano, J. McKinstry, C. Di Nolfo, T. Nayak, A. Andreopoulos, G. Garreau, M. Mendoza, J. Kusnitz, M. Debole, S. Esser, T. Delbruck, M. Flickner, D. Modha,
A Low Power, Fully Event-Based Gesture Recognition System,
{\it 2017 IEEE Conference on Computer Vision and Pattern Recognition} {\bf2017}.

\bibitem{Andreopoulos2018}
A. Andreopoulos, H. J. Kashyap, T. K. Nayak, A. Amir,  M. D. Flickner,
A Low Power, High Throughput, Fully Event-Based Stereo System,
{\it The IEEE Conference on Computer Vision and Pattern Recognition (CVPR)} {\bf2018}, pp. 7532-7542.

\bibitem{Ambrogio2018}
S. Ambrogio, P. Narayanan, H. Tsai, R. M. Shelby, I. Boybat, C. di Nolfo, S. Sidler, M. Giordano, M. Bodini, N. C. P. Farinha, B. Killeen, C. Cheng, Y. Jaoudi, G. W. Burr,
Equivalent-accuracy accelerated neural-network training using analogue memory,
{\it Nature} {\bf2018}, {\it558}, 60.

\bibitem{Churchland2004}
P. M. Churchland, Functionalism at Forty: A Critical Retrospective, 
 {\it J. Philosophy} {\bf2005}, {\it102}, 33.
 
\bibitem{SEP2016} 
Stanford Encyclopedia of Philosophy;  https://plato.stanford.edu/entries/dualism/ 

\bibitem{Dennett1996}
D. C. Dennett, Quantum Incoherence,  {\it Nature} {\bf1996}, {\it381}, 486.
Review of A. G. Cairns-Smith, Evolving the Mind: on the nature of matter and the origin of consciousness, Cambridge Univ. Press, 1996.

\bibitem{Chorost2013}
M. Chorost, Where Thomas Nagel Went Wrong, {\it The Chronicle of Higher Education} {\bf2013}, May 13.

\bibitem{Papineau2017}
D. Papineau, Competence without comprehension: The peculiar philosophical assumptions of Daniel Dennett,
{\it The Times Literary Supplement} {\bf2017}, June 28.

\bibitem{Landauer1991}
R. Landauer, Information is Physical, {\it Physics Today} {\bf1991}, {\it44}, 23.

\bibitem{Carhart-Harris2014}
R. L. Carhart-Harris, R. Leech, P. J. Hellyer, M. Shanahan, A. Feilding, E. Tagliazucchi, D. R.Chialvo, D. Nutt, The entropic brain: a theory of conscious states informed by neuroimaging research with psychedelic drugs, {\it Frontiers in Human Neuroscience} {\bf2014}, {\it8},  1. 
 
\bibitem{Carhart-Harris2017a}
R. L. Carhart-Harris, L. Roseman, M. Bolstridge, L. Demetriou, J. N. Pannekoek, M. B. Wall, M. Tanner, M. Kaelen, J. McGonigle, K. Murphy, R. Leech, H. V. Curran, D. J. Nutt, Psilocybin for treatment-resistant depression: fMRI-measured brain mechanisms, {\it Sci. Rep.} {\bf2017},  {\it7}, 13187.
 
\bibitem{Carhart-Harris2017b} 
R. L. Carhart-Harris,  D. J. Nutt, Serotonin and brain function: a tale of two receptors, {\it Journal of Psychopharmacology} {\bf2017}, {\it731},1091.

\bibitem{Fontes2015}
F. Palhano-Fontes, K. C. Andrade, L. F. Tofoli, A. C. Santos, J. A. S. Crippa, J. E. C. Hallak, S.  Ribeiro,  D. B. de Araujo,
The psychedelic state induced by ayahuasca modulates the activity and connectivity of the default mode network,
 {\it PLoS ONE}  {\bf2015},  {\it10},  e0118143. 

\bibitem{McKenna1994}
T. M. McKenna, T. A. McMullen, M. F. Shlesinger, The brain as a dynamic physical system, Neuroscience 
{\it Neuroscience} {\bf1994}, {\it60}, 587.

\bibitem{WattsStrogatz1998}
D. J. Watts, S. H. Strogatz, Collective dynamics of  "small-world" networks,  {\it Nature} {\bf1998}, {\it393}, 440.

\bibitem{Meunier2010}
D. Meunier, R. Lambiotte,  E. T. Bullmore, Modular hierarchically modular organization of brain networks, {\it Front. Neurosci.} {\bf2010}, {\it4}, 200.

\bibitem{BresslerMenon2010}
S. L. Bressler,  V. Menon,
Large-scale brain networks in cognition: emerging methods and principles,
{\it Trends  Cognit. Sci.} {\bf2010}, {\it14}, 277.

\bibitem{Menon2011}
V. Menon, Large-scale brain networks and psychopathology: a unifying triple network model,
{\it Trends  Cognit. Sci.} {\bf2011}, {\it15}, 483.

\bibitem{Andrews-Hanna_2012}
J. R. Andrews-Hanna, The Brain's Default Network and Its Adaptive Role in Internal Mentation, 
{\it The Neuroscientist} {\bf2012}, {\it18}, 251.

\bibitem{Andrews-Hanna_2014}
J. R. Andrews-Hanna, J. Smallwood,  R. N. Spreng,
The default network and self-generated thought: component processes, dynamic control, and clinical relevance,
{\it Ann. N.Y. Acad. Sci.} {\bf2014}, {\it1316}, 29.

\bibitem{Lee2017}
W. H. Lee and S. Frangou, 
Linking functional connectivity and dynamic properties of resting-state networks,
{\it Sci. Rep.} {\bf2017}, {\it7}, 16610

\bibitem{Qin2015}
 P. Qin, X. Wu, Z. Huang,  N. W. Duncan, W. Tang, A. Wolff, J. Hu, L. Gao, Y. Jin, X. Wu, J. Zhang, L. Lu, C. Wu, X. Qu, Y. Mao, X. Weng, J. Zhang,  G. Northoff, 
 How are different neural networks related to consciousness? 
 {\it Ann Neurol.}  {\bf2015}, {\it78}, 594. 
 
 \bibitem{Marek2018}
S. Marek, J. S.Siegel, Evan M.Gordon, R. V. Raut, C. Gratton, D. J. Newbold, M. Ortega, T. O. Laumann, B. Adeyemo, D. B.Miller, A. Zheng, K. C.Lopez, J. J. Berg, R. S.Coalson, A. L. Nguyen, D. Dierker, A. N.Van, C. R. Hoyt, N. U. F. Dosenbach,                                          
Spatial and Temporal Organization of the Individual Human Cerebellum,
{\it Neuron} {\bf2018}, {\it100}, 977.e7.

\bibitem{Berman2016} 
R. A. Berman, S. J. Gotts, H. M. McAdams, D. Greenstein, F. Lalonde, L. Clasen, R. E. Watsky, L. Shora, A. E. Ordonez, A. Raznahan, A. Martin, N. Gogtay, J. Rapoport,
Disrupted sensorimotor and social-cognitive networks underlie symptoms in childhood-onset schizophrenia,
{\it BRAIN} {\bf2016}, {\it139}, 276

\bibitem{Lant2016}
N. D. Lant, L. E. Gonzalez-Lara, A. M. Owen, Davinia Fernández-Espejo,
Relationship between the anterior forebrain mesocircuit and the default mode network in the structural bases of disorders of consciousness, 
{\it NeuroImage: Clinical} {\bf2016}, {\it10}, 27

\bibitem{Martino2016}
M. Martino, P. Magioncalda, Z. Huang, B. Conio, N. Piaggio, N. W. Duncan, G. Rocchi, A. Escelsior, V. Marozzi, A. Wolff, M. Inglese, M. Amore, G. Northoff, 
Contrasting variability patterns in the default mode and sensorimotor networks balance in bipolar depression and mania,
{\it PNAS}  {\bf2016}, {\it113}, 4824

\bibitem{Baars1988}
B. J. Baars,  {\it A cognitive theory of consciousness}, Cambridge University Press, New York {\bf1988}.

\bibitem{Baars2002}
B. J. Baars, The conscious access hypothesis: Origins and recent evidence.  {\it Trends in Cognitive Science} {\bf2002}, {\it6},  47.

\bibitem{Baars2005}
B. Baars, Global workspace theory of consciousness: Towards a cognitive
neuroscience of human experience? 
{\it Progress in Brain Research} {\bf2005}, {\it150}, 45.

\bibitem{Breakspear2017}
M. Breakspear, Dynamic models of large-scale brain activity,
{\it Nature Neuroscience} {\bf2017}, {\it20}, 340.

\bibitem{Olbrich2011}
E. Olbrich, P. Achermann, T. Wennekers, The sleeping brain as a complex system, {\it Phil. Trans. R. Soc. A} {\bf2011}, {\it369}, 3697.

\bibitem{Werner2007}
G. Werner, Metastability, Criticality and Phase Transitions in brain and its models, {\it Biosystems} {\bf2007}, {\it90}, 496.

\bibitem{Kitzbichler2009}
M. G. Kitzbichler, M. L. Smith, S. R. Christensen,  E. Bullmore, Broadband Criticality of Human Brain Network Synchronization. {\it  PLoS Comput. Biol.} {\bf2009}, {\it5},  e1000314.

\bibitem{Kitzbichler2015}
M. G. Kitzbichler,  E. T. Bullmore, Power Law Scaling in Human and Empty Room MEG Recordings, {\it PLoS Comput. Biol.} {\bf2015}, {\it11}, e1004175.

\bibitem{BeggsTimme2012}
J. M.Beggs,  N. Timme, Being critical of criticality in the brain,   {\it Frontiers in Physiology} {\bf2012}, {\it3}, 163.

\bibitem{Zhou2015}
D. W. Zhou, D. D. Mowrey, P. Tang,  Y. Xu,
Percolation Model of Sensory Transmission and Loss of Consciousness Under General Anesthesia,
{\it Phys. Rev. Lett.} {\bf2015}, {\it115}, 108103.

\bibitem{Papo2017}
D. Papo, J. Goni,  J. M. Buldu, Editorial: On the relation of dynamics and structure in brain networks, 
 {\it CHAOS} {\bf 2017}, {\it27}, 047201.

\bibitem{Racz2018}
F. S. Racz, Peter Mukli, Zoltan Nagy,  Andras Eke,
Multifractal dynamics of resting-state functional connectivity in the prefrontal cortex
{\it Physiol. Meas.} {\bf2018}, {\it39}, 024003.

\bibitem{Werner2010}
G. Werner, Fractals in the nervous system: conceptual implications for theoretical neuroscience, 
 {\it Frontiers in Physiology} {\bf2010}, {\it1}, 15.

\bibitem{Newman2006}
M. E. J. Newman, Modularity and community structure in networks,  {\it PNAS} {\bf2006}, {\it103}, 8577.

\bibitem{Enel2016}
P. Enel, E. Procyk, R. Quilodran, P. F. Dominey,
Reservoir Computing Properties of Neural Dynamics in Prefrontal Cortex,
{\it PLoS Comput. Biol.} {\bf2016}, {\it12}, e1004967.

\bibitem{Rodriguez2017}
N. Rodriguez, E. Izquierdo, Y.-Y. Ahn,
Optimal modularity and memory capacity of neural networks, Network Neuroscience (MIT) {\bf2019};  https://doi.org/10.1162/netn\_a\_00082; arXiv:1706.06511v2.

\bibitem{Taylor2017}
P. N. Taylor, Y. Wang,  M. Kaiser,
Within brain area tractography suggests local modularity using high resolution connectomics,
{\it Sci. Rep.} {\bf2017}, {\it7}, 39859.

\bibitem{Grossberg2017}
S. Grossberg,
Towards solving the hard problem of consciousness: The varieties of brain resonances and the conscious experiences that they support,
{\it Neural Networks} {\bf2017}, {\it87}, 38.

\bibitem{Seguin2018}
C. Seguin, Martijn P. van den Heuvel,  A. Zalesky, 
Navigation of brain networks, 
{\it PNAS} {\bf2018}, {\it115}, 6297.

\bibitem{vandenHeuvel2014}
M. P. van den Heuvel, A. Fornito,
Brain Networks in Schizophrenia,
{\it Neuropsychol Rev.} { \bf2014}, {\it24} 32.

\bibitem{Witter2015}
L. Witter, C. I. De Zeeuw,
Regional functionality of the cerebellum,
 {\it Current Opinion in Neurobiology} {\bf2015}, {\it33}, 150. 

\bibitem{Wagner2017}
M. J. Wagner, T. H. Kim, J. Savall, M. J. Schnitzer,  L. Luo,
Cerebellar granule cells encode the expectation of reward,
{\it Nature} {\bf2017}, {\it544}, 96.

\bibitem{Tognoli2014}
E. Tognoli,  J. A. Scott Kelso, 
The Metastable Brain,
{\it Neuron} {\bf2014}, {\it81}, 35.

\bibitem{Rossert2015}
C. R\"ossert, P. Dean,  J. Porrill, 
At the Edge of Chaos: How Cerebellar Granular Layer Network Dynamics Can Provide the Basis for Temporal Filters, PLoS {\it PLoS Comput. Biol,} {\bf2015}, {\it11},  e1004515.

\bibitem{Stepp2015}
N. Stepp, D. Plenz,  N. Srinivasa,
Synaptic Plasticity Enables Adaptive Self-Tuning Critical Networks,
{\it PLoS Comput. Biol.} {\bf2015}, {\it11}, e1004043.

\bibitem{Kauffman1991}
J. S. A. Kauffman, S. Johnsen, 
Coevolution to the edge of chaos: Coupled fitness landscapes, poised states, and coevolutionary avalanches,
{\it J. Theor. Biology} {\bf1991}, {\it149}, 467.

\bibitem{Watkins2016}
N. W. Watkins, G. Pruessner, S. C. Chapman, N. B. Crosby,  H. J. Jensen,
25 Years of Self-organized Criticality: Concepts and Controversies, {\it Space Sci. Rev.} {\bf2016}, {\it198}, 3.

\bibitem{Roli2018}
A. Roli, M. Villani, A. Filisetti,  R. Serra,
Dynamical Criticality: Overview and Open Questions, {\it J. Syst. Sci. Complex.} {\bf2018}, {\it31}, 647.

\bibitem{Schrauwen2008}
B. Schrauwen, L. Busing, R. Legenstein,
On Computational Power and the Order-Chaos Phase Transition in Reservoir Computing,
{\it Advances in Neural Information Processing Systems t21 (NIPS 2008)} {\bf2008}.

\bibitem{Jaeger2001}
H. Jaeger, The "echo state" approach to analysing and training recurrent neural networks.  {\it GMD Report 148, German National Research Institute for Computer Science} {\bf2001}.

\bibitem{LukoseviciusJaeger2009}
M. Lukosevicius, H. Jaeger, Reservoir computing approaches to recurrent neural network trainings, {\it Computer Science Review} {\bf2009}, {\it3}, 127.

\bibitem{Maass2009}
W. Maass, W. Liquid state machines: motivation, theory, and applications, Computability in context: computation and logic in the real world.
in {\it Computability in context: computation and logic in the real world,} (Eds: B. Cooper, A. Sorbi), World Scientific {\bf2009}, pp. 275-296. 

\bibitem{Yamazaki2007}
T. Yamazaki, S. Tanaka,
The cerebellum as a liquid state machine,
{\it Neural Networks} {\bf2007}, {\it20} 290.

\bibitem{Konkoli2018a}
Z. Konkoli, On developing theory of reservoir computing for sensing applications: the state weaving environment echo tracker (SWEET) algorithm,  {\it Int. J. Parallel, Emergent Distrib.}  {\bf2018}, {\it33}, 121.

\bibitem{Konkoli2018b}
Z. Konkoli, S. Nichele, M. Dale, S. Stepney, Reservoir Computing with Computational Matter, in {\it Computational Matter}, (Eds: S. Stepney, S. Rasmussen, M. Amos), Springer {\bf 2018}, pp. 269-293.

\bibitem{Inubushi2017}
M. Inubushi,  K. Yoshimura, Reservoir Computing Beyond Memory-Nonlinearity Trade-off,
{\it Sci. Rep.} {\bf2017}, {\it7}, 1019.

\bibitem{Sillin2013}
H. O. Sillin, R. Aguilera, H.-H. Shieh, A. V. Avizienis, M. Aono, A. Z. Stieg,  J. K. Gimzewski,
A theoretical and experimental study of neuromorphic atomic switch networks for reservoir computing, {\it Nanotechnology} {\bf2013}, {\it24}, 3840.

\bibitem{Demis2015}
E. C. Demis, R. Aguilera, H. O. Sillin, K. Scharnhorst, E. J. Sandouk, M. Aono, A. Z. Stieg,  J. K. Gimzewski,
Atomic switch networks - nanoarchitectonic design of a complex system for natural computing, {\it Nanotechnology} {\bf2015}, {\it26}, 204003.

\bibitem{VanderSande2017}
G. Van der Sande, D. Brunner, M. C. Soriano
Advances in photonic reservoir computing,
{\it Nanophotonics} {\bf2017}, {\it6}, 561.

\bibitem{Du2017}
C. Du, F. Cai, M. A. Zidan, W. Ma, S. Hwan Lee,  W. D. Lu,
Reservoir computing using dynamic memristors for temporal information processing, 
{\it Nature Commun.} {\bf2017}, {\it8}, 2204.

\bibitem{Pecqueur2018}
S. Pecqueur, M. Mastropasqua Talamo, David Gu\'erin, P. Blanchard, J. Roncali, D. Vuillaume,  F. Alibart,
Neuromorphic Time-Dependent Pattern Classification with Organic Electrochemical Transistor Arrays, {\it Adv. Electron. Mater.} {\bf 2018}, {\it4}, 1800166.

\bibitem{Fusi2016}
S. Fusi, E. K. Miller,  M. Rigotti,
Why neurons mix: high dimensionality for higher cognition,
{\it Current Opinion in Neurobiology} {\bf2016}, {\it37}, 66.

\bibitem{Fink2016}
S. B.Fink,
A Deeper Look at the "Neural Correlate of Consciousness",
{\it Front. Psychol. } {\bf2016}, {\it7}, 1044.

\bibitem{Tegmark2015}
 M. Tegmark, Consciousness as a state of matter,  {\it Chaos, Solitons \& Fractals} {\bf2015}, {\it76}, 238..
 
 \bibitem{Bestmann2013}
S. Bestmann,  E. Feredoes,
Combined neurostimulation and neuroimaging in cognitive neuroscience: past, present, and future,
{\it Ann. N.Y. Acad. Sci } {\bf2013}, {\it1296}, 11.

\bibitem{Sarasso2014}
S. Sarasso, M. Rosanova, A. G. Casali, S. Casarotto, M. Fecchio, M. Boly, O. Gosseries, G. Tononi, S. Laureys,  M. Massimini,
Quantifying Cortical EEG Responses to TMS in (Un)consciousness,
{\it Clin. EEG Neurosci.} {\bf2014}, {\it5}, 40.

\bibitem{Hallam2016}
G. P. Hallam, C. Whitney, M. Hymers, A. D. Gouws,  E. Jefferies,
Charting the effects of TMS with fMRI: Modulation of cortical recruitment within the distributed network supporting semantic control,
{\it Neuropsychologia} {\bf2016}, {\it93}, 40.

\bibitem{Hawco2017}
C. Hawco, J. L. Armony, Z. J. Daskalakis, M. T. Berlim, M. M. Chakravarty, G. B. Pike,  M. Lepage,
Differing Time of Onset of Concurrent TMS-fMRI during Associative Memory Encoding: A Measure of Dynamic Connectivity,
{\it Front. Hum. Neurosci.} {\bf2017}, {\it11}, 404.

\bibitem{Saari2018}
J. Saari, E. Kallioniemi, M. Tarvainen,  P. Julkunen,
Oscillatory TMS-EEG-Responses as a Measure of the Cortical Excitability Threshold, 
{\it IEEE Trans. Neural Syst. Rehab. Eng.} {\bf2018}, {\it26}, 383.

\bibitem{Leitao2017}
J. Leitão, A. Thielscher, J. Tuennerhoff, U. Noppeney,
Comparing TMS perturbations to occipital and parietal cortices in concurrent TMS-fMRI studies: Methodological considerations, 
{\it PLoS ONE} {\bf2017}, {\it12}, e0181438. 

\bibitem{Maguire2016}
P. Maguire, P. Moser,  R. Maguire,
Understanding Consciousness as Data Compression,
{\it Journal of Cognitive Science} {\bf2016} {\it17}, 63.

\bibitem{Maguire2014}
P. Maguire, P. Moser, R. Maguire,  V. Griffith, 
Is consciousness computable? Quantifying integrated information using algorithmic information theory, in {\it Proceedings of the 36th Annual Conference of the Cognitive Science Society, 2615-2620. Austin, TX},  (Eds. P. Bello, M. Guarini, M. McShane, B. Scassellati), Cognitive Science Society, {\bf2014}.

\bibitem{Aaronson2014}
S. Aaronson, Why I Am Not An Integrated Information Theorist (or, The Unconscious Expander), {\it Shtetl-Optimized} {\bf2014}; https://www.scottaaronson.com/blog/?p=1799.

\bibitem{Tononi2014}
G. Tononi {\bf2014}; https://www.scottaaronson.com/blog/?p=1823

\bibitem{Cerullo2015}
M. A. Cerullo, The Problem with Phi: A Critique of Integrated Information Theory,
{\it PloS Comput. Biol.}  {\bf2015}, {\it11},  e1004286.

\bibitem{Fallon2016}
F. Fallon, Integrated Information Theory of Consciousness,
{\it Internet Encyclopedia of Philosophy} {\bf2016};
http://www.iep.utm.edu/int-info/.

\bibitem{Fallon2018}
F. Fallon, Integrated Information, {\it The Routledge Handbook of Consciousness} (Ed. R. J. Gennaro), Taylor and Francis {\bf2018}.

\bibitem{Bayne2018}
T. Bayne,
On the axiomatic foundations of the integrated information theory of consciousness,
{\it Neuroscience of Consciousness} {\bf2018},  {\it4}, niy007.

\bibitem{Hilger2017}
K. Hilger, M. Ekman, C. J. Fiebach, U. Basten,
Intelligence is associated with the modular structure of intrinsic brain networks,
{\it Sci. Rep.} {\bf2017}, {\it7}, 16088.

 \bibitem{Murguialday2011}
 A. Ramos Murguialday, J. Hill, M. Bensch, S. Martens, S. Halder, F. Nijboer, B. Schoelkopf, N. Birbaumer, A. Gharabaghi,
Transition from the locked in to the completely locked-in state: A physiological analysis,
{\it Clinical Neurophysiology} {\bf2011}, {\it122},  925.

 \bibitem{Chaudhary2017}
U. Chaudhary, B. Xia, S. Silvoni, L. G. Cohen, N. Birbaumer,
Brain-Computer Interface-Based Communication in the Completely Locked-In State,
{\it PLoS Biol} {\bf2017}, {\it15}, e1002593.

 \bibitem{Guger2017}
C. Guger, R. Spataro, B. Z. Allison, A. Heilinger, R. Ortner, W. Cho, V. La Bella,
Complete Locked-in and Locked-in Patients: Command Following Assessment and Communication with Vibro-Tactile P300 and Motor Imagery Brain-Computer Interface Tools,
{\it Front. Neurosci.} {\bf2017} {\it11}, 251.

\bibitem{Lesenfants2018}
D. Lesenfants, D. Habbal, C. Chatelle, A. Soddu, S. Laureys, Q. Noirhomme,
Toward an Attention-Based Diagnostic Tool for Patients With Locked-in Syndrome,
{\it Clinical EEG and Neuroscience} {\bf2018}, {\it49}, 122. 

\bibitem{Rohaut2017}
B. Rohaut, F. Raimondo, D. Galanaud, M. Valente, J. D. Sitt, L. Naccache, Probing consciousness in a sensory-disconnected
paralyzed patient, {\it Brain Injury} {\bf2017}, {\it31}, 1398.
 
\bibitem{Espejo2015}
D. Espejo, S. Rossit, A. M. Owen, 
A Thalamocortical Mechanism for the Absence of Overt Motor Behavior in Covertly Aware Patients,
{\it JAMA Neurol.} {\bf2015}, {\it72}, 1442.

\bibitem{Reichle2001}
M. E. Raichle, A. M. MacLeod, A. Z. Snyder, W. J. Powers, D. A. Gusnard,  G. L. Shulman,
A default mode of brain function,
 {\it PNAS}, {\bf2001}, {\it98}, 676.

\bibitem{Reichle2007}
 M. E. Raichle,  A. Z. Snyder,
 A default mode of brain function: A brief history of an evolving idea,
{\it  NeuroImage}, {\bf2007}, {\it37}, 1083.

\bibitem{Hausknecht2017}
M. Hausknecht, W.-K. Li, M. Mauk,  P. Stone,
Machine Learning Capabilities of a Simulated Cerebellum,
{\it IEEE Trans. Neural Networks and Leaning Systems} {\bf2017},  {\it28}, 510.

\bibitem{Clausi2017}
S. Clausi, C. Iacobacci, M. Lupo, Giusy Olivito, M. Molinari, M. Leggio, The Role of the Cerebellum in Unconscious and Conscious Processing of Emotions: A Review,
{\it Appl. Sci.}, {\bf2017},  {\it7}, 521.

\bibitem{Fahrenfort2017}
J. J. Fahrenfort, J. van Leeuwen, C. N. L. Olivers,  H. Hogendoorn, Perceptual integration without conscious access, 
{\it PNAS} {\bf2017}, {\it114}, 3744. 

\bibitem{Roy2017}
D. S. Roy,T. Kitamura, T. Okuyama, S. K. Ogawa, C. Sun, Y. Obata, A. Yoshiki, S. Tonegawa,
Distinct Neural Circuits for the Formation and Retrieval of Episodic Memories,
{\it Cell} {\bf2017}, {\it170}, 1000.

\bibitem{Thompson2018}
E. H. Thompson, K. Kinden Lensjø, M. Brænne Wigestrand, A. Malthe-Sørenssen, T. Hafting, M. Fyhn,
Removal of perineuronal nets disrupts recall of a remote fear memory,
{\it PNAS}  {\bf2018}, {\it115}, 607.
 
\bibitem{Sergent2017}
C. Sergent, F. Faugeras, B. Rohaut, F. Perrin, M. Valente, C. Tallon-Baudry, L. Cohen, L. Naccache,
Multidimensional cognitive evaluation of patients with disorders of consciousness using EEG: A proof of concept study,
{\it NeuroImage: Clinical } {\bf2017}, {\it13}, 455.

\bibitem{Naccache2018}
L. Naccache,
Minimally conscious state or cortically mediated state?
{\it BRAIN} {\bf2018}, {\it141}, 949.

\bibitem{Bayne2016}
T. Bayne, J. Hohwy,  A. M. Owen,
Are There Levels of Consciousness?
{\it Ann. Neurol.}  {\bf2017}, {\it82}, 866.

\bibitem{Bayne2017}
T. Bayne, J. Hohwy, and A. M. Owen.
Reforming the Taxonomy in Disorders of Consciousness,
{\it Ann. Neurol.}  {\bf2017}, {\it82}, 866.

\bibitem{Silver2017}
D. Silver, J. Schrittwieser, K. Simonyan, I. Antonoglou, A. Huang, A. Guez, T. Hubert, L. Baker, M. Lai, A. Bolton, Y. Chen, T. Lillicrap, F. Hui, L. Sifre, G. van den Driessche, T. Graepel, D. Hassabis, 
Mastering the game of Go without human knowledge,
{\it Nature} {\bf2017}, {\it550}, 354.

\bibitem{VanDamme2017}
P. Van Damme, W. Robberecht, L. Van Den Bosch,
Modelling amyotrophic lateral sclerosis: progress and possibilities,
{\it Disease Models \& Mechanisms} {\bf2017}, {\it10}, 537.

\bibitem{List2017}
A. List, M. D. Rosenberg, A. Sherman,  M. Esterman,
Pattern classification of EEG signals reveals perceptual and attentional states,
{\it PLoS ONE} {\bf2017}, {\it12}, e0176349.

\bibitem{Deroy2016}
O. Deroy, C. Spence,  U. Noppeney,
Metacognition in Multisensory Perception,
{\it Trends in Cognitive Sciences} {\bf2016}, {\it20}, 736..

\bibitem{Deroy2014}
O. Deroy, Y.-C. Chen,  C. Spence,
Multisensory constraints on awareness, 
{\it Phil. Trans. R. Soc. B} {\bf2014}, {\it369}, 20130207.

\bibitem{OCallaghan2017}
C. O'Callaghan,
Grades of Multisensory Awareness, 
{\it Mind \& Language} {\bf2017}, {\it32}, 155.

\bibitem{Briscoe2017}
R. E. Briscoe,
Multisensory processing and perceptual consciousness: Part II, 
{\it Philosophy Compass} {\bf2017}, {\it12}, e12423. 

\bibitem{Liang2015}
Z. Liang, Y. Wang, X. Sun, D. Li, L. J. Voss, J. W. Sleigh, S. Hagihira,  X. Li,
EEG entropy measures in anesthesia,
{\it Front. Comput. Neurosci.} {\bf2015}, {\it9},16. 

\bibitem{Wirsich2018}
J. Wirsich, A.-L. Giraud,  S. Sadaghiani,
Concurrent EEG- and fMRI-derived functional connectomes exhibit linked dynamics,  {\bf2018}, 
bioRxiv (Nov. 7, 2018); doi: http://dx.doi.org/10.1101/464438 

\bibitem{Vidaurre2018}
D. Vidaurre, R. Abeysuriya, R. Becker, A. J. Quinn, F. Alfaro-Almagro, S. M. Smith,  M. W. Woolrich,
Discovering dynamic brain networks from big data in rest and task,
{\it NeuroImage} {\bf2018}, {\it180},  646.

\bibitem{Oken2014}
B. S. Oken, U. Orhan,  B. Roark, D. Erdogmus, A. Fowler, A. Mooney, B. Peters, M. Miller, M. B. Fried-Oken, 
Brain-computer interface with language model-EEG fusion for locked-in syndrome,
{\it Neurorehabil. Neural Repair.} {\bf2014},  {\it28}, 387.

\bibitem{Tian2017}
Y. Tian, H. Zhang, W. Xu, H. Zhang, L. Yang, S. Zheng, Y. Shi,
Spectral Entropy Can Predict Changes of Working Memory Performance Reduced by Short-Time Training in the Delayed-Match-to-Sample Task,
{\it Front. Hum. Neurosci.} {\bf2017}, {\it11}, 437.

\bibitem{Mateos2017}
D. M. Mateos, R. Wennberg, R. Guevara,  J. L. Perez Velazquez,
Consciousness as a global property of brain dynamic activity,
{\it Phys. Rev. E} {\bf2017}, {\it96}, 062410.

\bibitem{Schirrmeister2017}
R. T. Schirrmeister, J. T.  Springenberg, L. D. J. Fiederer,  M. Glasstetter, K. Eggensperger, M. Tangermann, F. Hutter, W. Burgard,  T. Ball,
Deep Learning With Convolutional Neural  Networks for EEG Decoding and Visualization,
{\it Human Brain Mapping} {\bf2017}, {\it38}, 5391.

\bibitem{Wen2018}
D. Wen, Z. Wei, Y. Zhou, G. Li, X. Zhang and W. Han,
Deep Learning Methods to Process fMRI Data and Their Application in the Diagnosis of Cognitive Impairment: A Brief Overview and Our Opinion,
{\it Front. Neuroinform.} {\bf 2018}, {\it12}, 23.

\bibitem{Hong2018}
K.-S. Hong, M. J. Khan, M. J. Hong,
Feature Extraction and Classification Methods for Hybrid fNIRS-EEG Brain-Computer Interfaces.
{\it Front. Hum. Neurosci.} {\bf2018}, {\it12}, 46.

\bibitem{Knoth2018}
I. S. Knoth, T. Lajnef, S. Rigoulot, K. Lacourse, P. Vannasing, J. L. Michaud, S. Jacquemont, P. Major, K. Jerbi,  S. Lipp\'e,
Auditory repetition suppression alterations in relation to cognitive functioning in fragile X syndrome: a combined EEG and machine learning approach,
{\it J. Neurodevelopmental Disorders}  {\bf2018}, {\it10}, 4.

\bibitem{Simpraga2017}
S. Simpraga, R. Alvarez-Jimenez, H. D. Mansvelder, J. M. A. van Gerven, G. Jan Groeneveld, S.-S. Poil, K. Linkenkaer-Hansen,
EEG machine learning for accurate detection of cholinergic intervention and Alzheimer's disease, 
{\it Sci. Rep.}  {\bf2017},  {\it7}, 5775.

\bibitem{Cukic2018}
M. \v{C}uki\'c, D. Pokrajac, M. Stoki\'c, S. Simi\'c, V. Radivojevi\'c,   M. Ljubisavljevi\'c,
EEG machine learning with Higuchi's fractal dimension and Sample Entropy as features  for successful detection of depression,  {\bf2018}; arXiv:1803.05985.

\bibitem{vanPutten2018}
M. J. A. M. van Putten, S. Olbrich,  M. Arns,
Predicting sex from brain rhythms with deep learning,
{\it Sci. Rep.}  {\bf2018}, {\it8}, 3069.

\bibitem{Shen2017}
G. Shen, T. Horikawa, K. Majima, Y. Kamitani, 
Deep image reconstruction from human brain activity, {\it PLoS Comput. Biol.}  {\bf2019}, {\it15}, e100663.

\bibitem{Shen2018}
G. Shen, K. Dwivedi, K. Majima, T. Horikawa, Y. Kamitani,
End-to-end deep image reconstruction from human brain activity, {\bf2018},  bioRxiv (2018).

 \bibitem{Fong2018}
R. C. Fong, W. J. Scheirer,  D. D. Cox,
Using human brain activity to guide machine learning,
{\it Sci. Rep.} {\bf2018}, {\it8}, 5397.

\bibitem{Vansteensel2016}
M. J. Vansteensel, E. G. M. Pels, M. G. Bleichner, M. P. Branco, T.  Denison, Z. V. Freudenburg, P. Gosselaar, S. Leinders, T. H. Ottens, M. A. Van Den Boom, P. C. Van Rijen, E. J. Aarnoutse, N.F. Ramsey,
Fully Implanted Brain-Computer Interface in a Locked-In Patient with ALS,
{\it N. Engl. J. Med.} {\bf2016}, {\it375}, 2060.

\bibitem{Wang2013}
Z. Wang, J. R. Busemeyer, H. Atmanspacher, E. M. Pothos,
The Potential of Using Quantum Theory to Build Models of Cognition, {\it Topics in Cognitive Sciences} {\bf2013}, {\it5}, 672. 

\bibitem{Lukasik2017}
A. \L ukasik, 
Quantum models of cognition and decision,
{\it Int. J. Parallel, Emergent and Distributed Systems} {\bf2018}, {\it33}, 3, 336.

\bibitem{Broekaert2017}
J. Broekaert, I. Basieva, P. Blasiak,  E. M. Pothos,
Quantum-like dynamics applied to cognition: a consideration of available options,
{\it Phil. Trans. R. Soc. A} {\bf2017}, {\it375}, 20160387.

\bibitem{Pothos2017}
E. M. Pothos, J. R. Busemeyer, R. M. Shiffrin,  J. M. Yearsley,
{\it J. Experimental Psychology: General} {\bf2017}, {\it146}, 968.

\bibitem{BarrosOas2017}
J. Acacio de Barros,  G. Oas,
Quantum Cognition, Neural Oscillators, and Negative Probabilities,
in {\it The Palgrave Handbook of Quantum Models in Social Science}, (Eds. E. Haven, A. Khrennikov), {\bf2017}.

\bibitem{Flitney2002}
A. P. Flitney,  D. Abbott, An introduction to quantum game theory, 
{\it Fluctuation and Noise Letters} {\bf2002}, {\it2}(4), R175.

\bibitem{BrunnerLinden2013}
N. Brunner,  N. Linden,
Connection between Bell nonlocality and Bayesian game theory,
{\it Nature Communications} {\bf2013} {\it4}, 2057.

\bibitem{Bang2016}
J. Bang, J. Ryu, M. Paw\l owski, B. S. Ham,  J. Lee,
Quantum-mechanical machinery for rational decision-making in classical guessing game, 
{\it Sci. Rep.} {\bf2016}, {\it6}, 21424.

\bibitem{BusemeyerBruza2012}
J. R. Busemeyer,  P. Bruza, {\it Quantum models of cognition and decision making}, Cambridge, UK, Cambridge University Press, {\bf2012}.

\bibitem{White2016}
L. C. White, E. M. Pothos,  J. R. Busemeyer,
Insights from quantum cognitive models for organizational decision making,
{\it J.  Applied Research in Memory and Cognition} {\bf2015}, {\it4}, 229.

\bibitem{Martinez2016}
I. Mart\'inez-Mart\'inez,  E. S\'anchez-Burillo,
Quantum stochastic walks on networks for decision-making,
{\it Sci. Rep.} {\bf2016}, {\it6}, 23812.

\bibitem{Baaquie2013}
B. E. Baaquie,
Financial modeling and quantum mathematics, 
{\it Computers and Mathematics with Applications} {\bf2013}, {\it65} 1665.

\bibitem{Jedlicka2017}
P. Jedlicka Revisiting the Quantum Brain Hypothesis: Toward Quantum (Neuro)biology?
{\it Front. Mol. Neurosci.} {\bf2017}, {\it10}, 366.

\bibitem{Bourget2004}
D. Bourget, Quantum Leaps in Philosophy of Mind,  {\it J. Consciousness Studies} {\bf2004}, {\it11}, 17.

\bibitem{Georgiev2015}
D. Georgiev, Mind Efforts, Quantum Zeno Effect and Environmental Decoherence,  {\it NeuroQuantology} {\bf2015}, {\it13}, 179.

\bibitem{Georgiev2018}
D. Georgiev, {\it Quantum Information and Consciousness a gentle introduction} {\bf2018}.

\bibitem{Fisher2017}
M. P. A. Fisher, Quantum cognition: The possibility of processing with nuclear spins in the brain,  {\it Annals of Physics} {\bf2015}, {\it362}, 593.

\bibitem{CaiPlenio2013}
J. Cai,  M. Plenio, Chemical Compass Model for Avian Magnetoreception as a Quantum Coherent Device, {\it Phys. Rev. Lett.} {\bf2013}, {\it111}, 230503.

\bibitem{Tiersch2014}
M. Tiersch, G. G. Guerreschi, J. Clausen,  H. J. Briegel, Approaches to Measuring Entanglement in Chemical Magnetometers, {\it J. Phys. Chem. A} {\bf2014}, {\it118}, 13.

\bibitem{Mohseni2014}
M. Mohseni, Y. Omar, G. S. Engel, M. B. Plenio,  {\it Quantum Effects in Biology}, Cambridge University Press,  {\bf2014}.

\bibitem{Schwartz2005}
J. M. Schwartz, H. P. Stapp,  M. Beauregard,
Quantum physics in neuroscience and psychology: a neurophysical model of mind-brain interaction,
{\it Phil. Trans. R. Soc. B} {\bf2005}, {\it60}, 1309. 

\bibitem{Stapp2011}
H. P. Stapp, {\it Mindful Universe: Quantum Mechanics and the Participating Observer}, 2nd edition, Springer, {\bf2011}.

\bibitem{Stapp2017}
H. P. Stapp, Retrocausation in quantum mechanics and the effects of minds on the creation of physical reality,  {\it AIP Conf. Proc} {\bf2017}, {\it1841}, 040001.

\bibitem{Hameroff1988} 
S. R. Hameroff, S. Rasmussen,  B. Mansson,  Molecular automata in microtubules: basic computational logic of the living state?, in {\it Artificial Life, SF/studies in the sciences of complexity} (Ed. C. Langton). AddisonWesley, New York, {\bf1988}..

\bibitem{Penrose1994}
R. Penrose, {\it Shadows of the Mind, A Search for the Missing Science of Consciousness}, Oxford University Press, {\bf1994}.

\bibitem{HameroffPenrose1996}
S. Hameroff, R. Penrose,
Orchestrated reduction of quantum coherence in brain microtubules: A model for consciousness,
{\it Mathematics and Computers in Simulation} {\bf1996}, {\it40}, 453. 

\bibitem{Tegmark2000}
M. Tegmark, The Importance of Quantum Decoherence in Brain Processes, {\it Phys. Rev. E} {\bf2000}, {\it1}, 4194.

\bibitem{Preskill2018}
J. Preskill,
Quantum Computing in the NISQ era and beyond, {\it Quantum}  {\bf2018}, {\it2}, 79.

\bibitem{HBP2018}
{\it The Human Brain Project (HBP} {\bf2018}; 
https://www.humanbrainproject.eu/en/brain-simulation/brain-simulation-platform/

\bibitem{Schuld2015}
M. Schuld, I. Sinayskiy, F. Petruccione,
Simulating a perceptron on a quantum computer,
{\it Phys. Lett. A} {\bf2015}, {\it379}, 660.

\bibitem{Wiebe2016}
N. Wiebe, A. Kapoor,  K. Svore,
Quantum Perceptron Models,
{\it 30th Conference on Neural Information Processing Systems (NIPS 2016)}, Barcelona, Spain.

\bibitem{Cao2017}
Y. Cao, G. G. Guerreschi, A. Aspuru-Guzik, Quantum Neuron: An elementary building block for machine learning on a quantum computer  {\bf2017}; arXiv:1711.11240.

\bibitem{Wan2017}
K. H. Wan, O. Dahlsten, H. Kristj\'ansson, R. Gardner,  M. S. Kim,
Quantum generalisation of feedforward neural networks,
{\it npj Quantum Information}, {\bf2017}, {\it3}, 36.

\bibitem{Pfeiffer2016}
P. Pfeiffer, I. L. Egusquiza, M. Di Ventra, M. Sanz, E. Solano, 
Quantum Memristor, 
{\it Sci. Rep.} {\bf2016}, {\it6,} 29507.

\bibitem{Salmilehto2017}
J. Salmilehto, F. Deppe, M. Di Ventra, M. Sanz,  E. Solano, 
Quantum Memristors with Superconducting Circuits, 
{\it Sci. Rep. } {\bf2017}, {\it7}, 42044.

\bibitem{Cheng2018}
X.-H. Cheng, T. Gonzalez-Raya, X. Chen, M. Sanz,  E. Solano, 
Quantized Hodgkin-Huxley Model for Quantum Neurons {\bf2018}; arXiv:1807.10698v1.

\bibitem{Silva2018}
F. Silva, M. Sanz, J. Seixas, E. Solano,  Y. Omar,
Perceptrons from Memristors {\bf2018}; arXiv:1807.04912v1.

\bibitem{Fujii2017}
K. Fujii,  K. Nakajima, Harnessing Disordered-Ensemble Quantum Dynamics for Machine Learning,  {\it Phys. Rev. Applied} {\bf2017}, {\it8}, 024030.

\bibitem{Nakajima2018}
K. Nakajima, K. Fuji, M. Negoro, K, Mitarai,  M. Kitagawa,
Boosting computational power through spatial multiplexing in quantum reservoir computing, {\it Phys. Rev. Applied}  {\bf2019}, {\it11}, 034021.

\bibitem{Kutvonen2018}
A. Kutvonen, K. Fujii,  T. Sagawa,
Recurrent neural networks running on quantum spins: memory accuracy and capacity {\bf2018}; arXiv:1807.03947v1. 

\bibitem{Biamonte2017}
J. Biamonte, P. Wittek, N. Pancotti, P. Rebentrost, N. Wiebe,  S. Lloyd, 
Quantum machine learning,
{\it Nature} {\bf2017}, {\it549}, 195.

\bibitem{Rebentrost2017}
P. Rebentrost, T. R. Bromley, C. Weedbrook,  S. Lloyd,
A Quantum Hopfield Neural Network,  Phys. Rev. A  {\bf2018}, {\it98,} 042308.

\bibitem{Rotondo2018}
P. Rotondo M. Marcuzzi, J. P. Garrahan, I. Lesanovsky,  M M\"uller, Open quantum generalisation of Hopfield neural networks, {\it J. Phys. A: Math. Theor.} {\bf2018}, {\it51}, 115301.

\bibitem{Amin2016}
M. H. Amin, E. Andriyash, J. Rolfe, B. Kulchytskyy, B. Melko, Quantum Boltzmann machine, {\bf2016}; arXiv:1601.02036.
    
\bibitem{Deng2017}
D.-L. Deng, X. Li,  S. Das Sarma,
Quantum Entanglement in Neural Network States,
{\it Phys. Rev. X} {\bf2017}, {\it7}, 021021.

\bibitem{Mitarai2018}
K. Mitarai, M. Negoro, M. Kitagawa,  K. Fujii,
Quantum circuit learning,
{\it Phys.  Rev. A} {\bf2018}, {\it98}, 032309.

\bibitem{FarhiNeven2018}
E. Farhi,  H. Neven, Classification with Quantum Neural Networks on Near Term Processors,  {\bf2018}; arXiv:1802.06002v1.

\bibitem{Briegel2012}
H. J. Briegel, On creative machines and the physical origins of free behavior, {\it Sci. Rep.} {\bf2012}, {\it2}, 522.

\bibitem{Mautner2015}
J. Mautner, A. Makmal, D. Manzano, M. Tiersch, H. J. Briegel, Projective simulation for classical learning agents: a comprehensive investigation, {\it New Gener. Comput.} {\bf2015}, {\it3}, 69.

\bibitem{Dunjko2015}
V. Dunjko, N. Friis, H. J. Briegel, Quantum-enhanced deliberation of learning agents using trapped ions, {\it New J. Phys. } {\bf2015}, {\it17}, 023006.

\bibitem{Dunjko2016}
V. Dunjko, J. M. Taylor, H. J. Briegel, Quantum-Enhanced Machine Learning, {\it Phys. Rev. Lett.} {\bf2016}, {\it117}, 130501.

\bibitem{Melnikov2017}
A. A. Melnikov, A. Makmal, V. Dunjko, H. J. Briegel,
Projective simulation with generalization, 
{\it Sci. Rep.} {\bf2017} {\it7}, 14430.

\bibitem{Melnikov2018}
A. A. Melnikov, H. Poulsen Nautrup, M. Krenn, V. Dunjko, M. Tiersch, A. Zeilinger, H. J. Briegel,
Active learning machine learns to create new quantum experiments,  {\it PNAS} {\bf2018}, {\it115}, 1221.

\bibitem{Yamamoto2017}
Y. Yamamoto, K. Aihara , T. Leleu, K.-i. Kawarabayashi, S. Kako, M. Fejer, K. Inoue,  H. Takesue,
Coherent Ising machines - optical neural networks operating at the quantum limit,
{\it npj Quantum Information} {\bf2017}, {\it49}.

\bibitem{Hamerly2018}
R. Hamerly, T. Inagaki, P. L. McMahon, D. Venturelli, A. Marandi, T. Onodera, E. Ng, C. Langrock, K. Inaba,
T. Honjo, K. Enbutsu, T. Umeki, R. Kasahara, S. Utsunomiya, S. Kako, K.-i. Kawarabayashi, R. L. Byer,
M. M. Fejer, H. Mabuchi, E. Rieffel, H. Takesue, Y. Yamamoto, Scaling advantages of all-to-all connectivity in physical annealers: the Coherent Ising Machine vs. D-Wave 2000Q, {\bf2018},  arXiv:1805.05217v1.

\bibitem{King2018}
A. D. King, W. Bernoudy, J. King, A. J. Berkley,  T. Lanting,
Emulating the coherent Ising machine with a mean-field algorithm, 
{\bf2018}; arXiv:1806.08422v1.

\bibitem{Snaprud2018}
P. Snaprud, The consciousness wager, {\it New Scientist}, {\bf2018},  23 June, p.23.


\end{thebibliography}
\end{document}